\documentclass[11pt,onecolumn]{article}

\usepackage{amsmath}
\usepackage{amssymb}
\usepackage{bm}
\usepackage{color}
 \usepackage{graphicx} 
\usepackage{lipsum} 
\usepackage{multirow}
\usepackage{overpic}
\usepackage{psfrag}
\usepackage{subeqnarray}
\usepackage{ulem} 
\usepackage[colorlinks=true]{hyperref} 

\def\xx{\mathbf{x}}%

\def\be{\begin{equation}}
\def\ee{\end{equation}}
\def\CC{\mathbf{C}}
\def\eee {\mathbf{e}}
\def\ff {\mathbf{f}}
\def\FF {\mathbf{F}}
\def\II{\mathbf{I}}
\def\LL{\mathbf{L}}
\def\nn{\mathbf{n}}
\def\NN{\mathbf{N}}

\def\ppsi{\boldsymbol{\psi}}
\def\RR{\mathbf{R}}
\def\SS{\mathbf{S}}
\def\uu {\mathbf{u}}
\def\UU {\mathbf{U}}

\def\Um{\overline{U}}
\def\Vm{\overline{V}}
\def\UUm{\overline{\mathbf{U}}}
\def\zzz {\mathbf{0}}
\def\feedback {\mathbf{fb}}
\def\dom    {{\mbox{I}}}
\def\subdom {{\mbox{J}}}
\def\totalvortvec{\boldsymbol{\Omega}}
\def\totalvortscal{\Omega}
\def\meanvort{\overline{\Omega}}
\def\Ubulk  {        U_\infty }
\def\UUbase {\mathbf{U}_b     }

\newcommand\Rey{\mbox{\textit{Re}}} 
\newcommand{\ps}   [2]{\ensuremath{\left(\left. {#1} \, \right| \, {#2} \right) }}

\providecommand\bdelta{\boldsymbol{\delta}}\providecommand\bnabla{\boldsymbol{\nabla}}
\providecommand\bcdot{\boldsymbol{\cdot}}
\providecommand\bcol{\boldsymbol{:}}

\begin{document}

\title{
Saturation of a turbulent mixing layer over a cavity:\\
response to harmonic forcing around mean flows
}

\author{
E. Boujo, M. Bauerheim and N. Noiray 
\\
\small CAPS Lab., Mechanical and Process Engineering Dept., ETHZ, CH-8092 Z\"urich, Switzerland
}

\date{\today}

\maketitle

\graphicspath{{./}}


\noindent 
Turbulent mixing layers over cavities can couple with acoustic waves and lead to undesired oscillations. To understand the nonlinear aspects of this phenomenon, a turbulent mixing layer over a deep cavity is considered and its response to harmonic forcing is analysed with large-eddy simulations (LES) and linearised Navier--Stokes equations (LNSE). The Reynolds number is $\Rey$=150~000. As a model of incoming acoustic perturbations, spatially uniform time-harmonic velocity forcing is applied at the cavity end, with amplitudes spanning the wide range 0.045--8.9$\%$ of the main channel bulk velocity. Compressible LES provide reference nonlinear responses of the shear layer, and the associated mean flows. Linear responses are calculated with the incompressible LNSE around the LES mean flows; they predict well the amplification (both measured with kinetic energy and with a proxy for vortex sound production in the mixing layer) and capture the nonlinear saturation observed as the forcing amplitude increases and the mixing layer thickens. Perhaps surprisingly, LNSE calculations  based on a monochromatic (single frequency) assumption yield a good agreement even though higher harmonics and their nonlinear interaction (Reynolds stresses) are not negligible. However, it is found that the leading Reynolds stresses do not force the mixing layer efficiently, as shown by a comparison with the optimal volume forcing obtained from a resolvent analysis. Therefore they cannot fully benefit from the potential for amplification available in the flow. Finally, the sensitivity of the optimal harmonic forcing at the cavity end is computed with an adjoint method. The sensitivities to mean flow modification and  to a localised feedback (structural sensitivity) both identify the upstream cavity corner as the region where a small-amplitude modification has the strongest effect. This can guide in a systematic way the design of strategies aiming at controlling the amplification and saturation mechanisms.

\bigskip


\medskip
\hrule

\vspace{1cm}

\section{Introduction}

Flow over a cavity leads to a variety of interesting phenomena, including radiated noise in the form of broadband and discrete components. 
Related applications are numerous in aeronautics (wheel wells), ground transportation (pantograph cavities, door gaps, open windows),
turbomachinery (bleed slots for secondary air supply in compressors) 
and other energy-related systems (T-junctions and side branches in pipe networks for air, water, steam or gas).
Therefore, it comes as no surprise that, for more than 60 years, many studies have investigated cavity flows.
In general, sustained vortical oscillations in the shear layer and acoustic oscillations result from a  process involving hydrodynamics and acoustics, although details depend on the specific configuration.
(See for instance reviews by \cite{Rockwell1978, Rockwell1979, Rockwell1983, Rowley2006, Tonon2011, Morris11}.)
In shallow cavities, vortical disturbances in the shear layer impinge on the downstream cavity corner, and cavity tones may be generated by a feedback mechanism (hydrodynamic/acoustic feedback in incompressible/compressible flows at small/large Mach number).
In deep cavities, 
resonant pipe tones may be generated if
 a cavity acoustic mode (standing wave) is excited
by the shear layer (turbulence-induced broadband  excitation and/or instability-induced narrowband excitation).

Many linear models have been developed for predicting oscillation frequencies (\cite{Rossiter1964, Tam1978, Kooijman2004, Alvarez2004}). However, accounting for nonlinear saturation and predicting oscillation amplitude remains a challenge. 
Full Navier--Stokes simulations are computationally expensive, especially in the turbulent regime, and simpler methods are still few.
\cite{Rowley2006} mention a model by \cite{Cain1996} which ``\textit{assumes oscillations at the Rossiter frequencies, and assumed nonlinearities enter through saturation of the shear layer: as the amplitude of oscillation grows, Reynolds stresses increase, and the shear layer spreads, decreasing the amplification rate of disturbances. The total amplification of a disturbance is computed around the loop, and an iterative procedure is used to converge to the final oscillation amplitude}''.
Because this procedure assumes specific nonlinearities, it may be too simplified to give accurate predictions in a wide range of conditions; however, its description of the saturation mechanism is particularly interesting because it probably captures the key ingredients at play.
It also points to a recent study by \cite{Mantic2016} who used a related description for  predicting the hydrodynamic response to harmonic forcing in the laminar flow over a backward-facing step. Their model is a system of two equations: the response to harmonic forcing at frequency $\omega_1$ is given by a linear equation (Navier--Stokes operator linearised around the mean flow), while nonlinear interaction of the response with itself modifies the mean flow. 
As nonlinear saturation effects increase, the mean flow and the linear response progressively converge (both in the iterative algorithm and in the physical flow) to a steady regime. 
In this semi-linear self-consistent model, the only assumption is that higher harmonics can be neglected altogether, i.e. they have no effect on the linear response at $\omega_1$, nor on the mean flow correction. 

In this paper we consider the flow over a deep cavity at large Reynolds number $\Rey$ and small Mach number $M$.
The flow can be seen as a system of two coupled elements: 
the incompressible shear layer (hydrodynamic element), and the compressible volume of fluid inside the cavity (acoustic element). 
In order to make a first step toward the simple prediction of oscillation amplitudes in aeroacoustic systems, we consider separately the shear layer and the deep cavity. 
In this study, we treat the cavity as an external element, and we focus specifically on the hydrodynamic response of the shear layer to a prescribed harmonic forcing.
This forcing is chosen as a plane wave coming from the cavity end, mimicking the dominant acoustic resonance mode (quarter-wave mode) at frequency $\omega_1$.
Motivated by the 
description mentioned earlier in terms of mean flow and harmonic fluctuations, we consider the linear response around the mean flow. 
In practice, the response to the prescribed forcing is obtained with the Linearised Navier-Stokes Equations (LNSE) incorporating a turbulence model, while the effect of higher harmonics on the response at $\omega_1$ is neglected.
We note that the wavelength of the observed cavity resonance mode is much larger than the shear layer width, such that one can make the  compactness assumption, neglect compressibility effects and use incompressible LNSE.
For simplicity, the mean flow is taken from nonlinear Navier-Stokes simulations carried out independently with Large-Eddy Simulations (LES) with harmonic forcing at various amplitudes, thus automatically taking into account the effect of higher harmonics on the mean flow.

Our study addresses several questions:
is it possible to predict accurately the response of the shear layer at different forcing amplitudes using LNSE around the mean flow? 
Can the  saturation mechanism be captured?
Can one neglect the effect of higher harmonics on the linear response?
None of these questions has obvious a priori answers.
In the laminar regime, linear stability analysis around mean flows has been shown to produce relevant results in some cases while failing in other cases. 
For instance, the  frequency of limit-cycle oscillations in the flow past a circular cylinder is well predicted by the dominant linear eigenvalue calculated around the mean flow (\cite{Barkley06}). This led \cite{Mantic2014} to build a self-consistent model for stability analysis in the same vein as that for harmonic response. 
\cite{Turton15} observes that linear stability analysis around the mean flow in a laminar thermosolutal convection system reproduces well the nonlinear characteristics of travelling waves; it fails, however, to produce meaningful results for standing waves. It was proposed that the reason might lie in the second harmonic being negligible for travelling waves, and non-negligible for standing waves. 
This is to be related to the earlier weakly nonlinear analysis of \cite{Sipp2007}, who formulated conditions on the second harmonic for the validity of stability analysis around mean flows, and presented a counterexample in a square open cavity.
Recently, \cite{Meliga2017JFM} extended the self-consistent model, incorporating the second harmonic.
In the turbulent regime, linear stability analysis and linear harmonic response calculations around mean flows are common, both in parallel and global settings (\cite{DelAlamo06, Piot06, Pujals09, Hwang10, Marquillie11, Meliga12D, Iungo13, Gikadi14, 
Mettot14, 
Oberleithner2015, Beneddine2016, Edstrand2016, Tammi2016}). 
It is not clear, however, if the structure and the amplitude of the response to harmonic forcing are meaningful in general, and if the saturation process can be captured accurately. This is what we assess for the deep cavity of this study.

The paper is organised as follows.
The configuration and the mean flow obtained from LES with different forcing amplitudes are presented in \S~\ref{sec:Mean}.
Section~\ref{sec:LinResp} is devoted to the mean-flow linear response calculated with the LNSE: the problem formulation and numerical method are detailed in \S\S~\ref{sec:Problem} and~\ref{sec:Numerical}, respectively. Results and comparison with LES results are given in \S\S~\ref{sec:ResultsUnforced}-\ref{sec:ResultsForced}.
Next, the effect of higher harmonics is investigated in \S~\ref{sec:ValidMonochrom}, in particular via consideration of optimal forcings (resolvent analysis).
Finally, \S~\ref{sec:Sensitivity} presents 
results from a sensitivity analysis that identifies  regions where a flow modification or a localised feedback   have the largest effect on the optimal harmonic response, which provides useful information for control design.
Conclusions are drawn in \S~\ref{sec:conclu}.

%
\section{Geometry and mean flow}
\label{sec:Mean}

\begin{figure}[] 
\centerline{
\includegraphics[height=6cm]{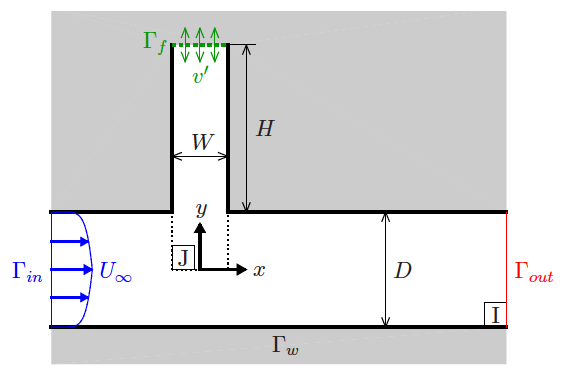} 
} 
\vspace{-0.2cm}
\caption{
Geometry and flow configuration (see main text for dimensions). 
LNSE calculations are performed in the 2D domain $\dom$, around mean flows from LES performed in a 3D domain (same 2D cross-section $\dom$, spanwise extension $L$). 
The subdomain $\subdom$ is used in  measures (\ref{eq:gain_KE_subdom}) and (\ref{eq:G_coriolis}) of the response.
} 
\label{fig:geom}
\end{figure}

We consider a straight rectangular channel of height $D=62$~mm, featuring on one side a deep cavity of width $W=30$~mm and depth $H=90$~mm (aspect ratio $W/H=0.33$).
Both channel and cavity are of 
spanwise extension $L=10$~mm.
A two-dimensional (2D) cross-section $\dom$ is shown in figure~\ref{fig:geom}.
The $x$, $y$ and $z$ direction are denoted streamwise, vertical and spanwise, respectively.
With an inlet bulk velocity $\Ubulk=56$~m/s and a speed of sound $c_0=340$~m/s, the air flow in this channel corresponds to a relatively small Mach number $M=\Ubulk/c_0=0.16$ and a large Reynolds number $\Rey = \Ubulk W/\nu = 1.5\times10^5$.

The effect of an acoustic forcing imposed at the cavity end is investigated by means of three-dimensional (3D) compressible large-eddy simulations (LES). 
At the inlet $\Gamma_{in}$ the incoming flow has a turbulent power-law profile of exponent 0.7.
Both inlet $\Gamma_{in}$ and outlet $\Gamma_{out}$ are acoustically non-reflecting.
A no-slip boundary condition is set on the walls $\Gamma_w$.
At the cavity end $\Gamma_{f}$, a vertical and spatially uniform, time-harmonic forcing $(0,v'\cos(\omega_1 t),0)$ is prescribed via a propagative acoustic wave with the NSCBC conditions (\cite{Poinsot92}).
In this study, the forcing frequency is set to $\omega_1/2\pi=750$~Hz, close to the frequency of a marginally stable eigenmode  and of the largest harmonic response in the unforced flow (see \S~\ref{sec:ResultsUnforced}).
More details about the numerical method are given in~\cite{Bauerheim17}.
The forcing amplitude $v'$ is varied  over more than two orders of magnitude, between $0.025$ and 5.0~m/s (relative velocity $v'/\Ubulk$ between $0.045\%$ and $8.9\%$).

\begin{figure}[] 
\centerline{
\includegraphics[height=12.7cm]{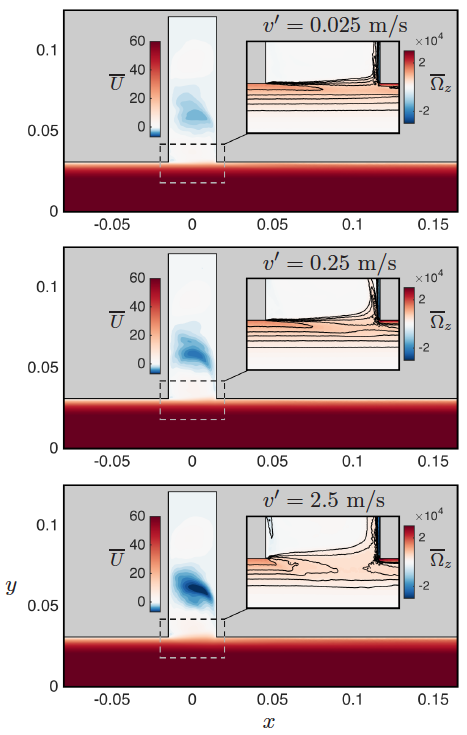}
}
\vspace{-0.2cm}
\caption{
Mean streamwise velocity $\Um$ from LES, at forcing amplitudes $v'=0.025$, $0.25$ and $2.5$~m/s.
Inset: close-up view of the mean spanwise vorticity $\meanvort_z = \partial_x \Vm - \partial_y \Um$ in the shear layer.
Main flow from left to right.
} 
\label{fig:vort}
\end{figure}

The LES mean flow obtained for different forcing amplitudes is shown figure~\ref{fig:vort}.
The streamwise velocity $\Um$ quickly decreases from $\Ubulk$ to 0 in the shear layer. A recirculation region is present inside the cavity, and becomes stronger with the forcing amplitude (maximum negative velocity between -3 and -9~m/s.)
As shown in the insets, the shear layer thickens and becomes weaker with $x$, and the mean spanwise vorticity $\meanvort_z = \partial_x \Vm - \partial_y \Um$ clearly diffuses.
Larger forcing amplitudes yield a thicker and weaker shear layer.

\section{Linear response of the mean flow to harmonic forcing}
\label{sec:LinResp}

\subsection{Problem formulation}
\label{sec:Problem}

We start from the Navier--Stokes equations
\begin{align}
\partial_t \UU + \NN(\UU)
= \FF,
\label{eq:NS}
\end{align}	
governing the dynamics of velocity $\UU(\xx,t)$ and pressure $P(\xx,t)$,
where 
$$\NN(\UU) = (\UU \bcdot \bnabla) \UU + \frac{1}{\rho} \bnabla P - \nu \bnabla^2 \UU$$
is the nonlinear incompressible Navier--Stokes operator
and $\rho$ the fluid density.
In this study we restrict our attention to incompressible flows since the Mach number is small, but the spirit of the derivation is similar for compressible flows.
Hereafter, we therefore omit the continuity equation $\bnabla\bcdot\UU=0$.
The term $\FF(\xx,t)$ denotes a space and time-dependent forcing applied either at a boundary or inside the domain.
Following \cite{Reynolds72_3}, the flow is decomposed into its time-averaged component $\UUm(\xx)$,  
coherent fluctuations $\widetilde\uu(\xx,t)$ 
and turbulent fluctuations $\uu'(\xx,t)$:
\begin{align}
 \UU  = \UUm  + \widetilde\uu + \uu'.
\label{eq:decomp}
\end{align}	
By construction, time averaging $\overline{\cdot}$ yields the steady mean flow $\UUm$ and removes all fluctuations 
($\overline{\widetilde\uu + \uu'}=\zzz$),
while phase averaging $\langle \cdot \rangle $
removes incoherent fluctuations
($\langle \UU \rangle = \UUm+\widetilde\uu$, $\langle \uu' \rangle=\zzz$).
(Similar notations are used for other quantities, e.g. pressure, forcing and spanwise vorticity.)
Substituting this decomposition 
into (\ref{eq:NS}) yields coupled equations for the mean flow and coherent fluctuations:
\begin{align}
\NN(\UUm) &= 
-\bnabla \bcdot \left( \overline{\widetilde\uu \widetilde\uu} + \overline{\uu' \uu'} \right)
+\overline{\FF} ,
\label{eq:mean}
\\
\partial_t \widetilde\uu + \LL(\UUm) \widetilde\uu 
&= 
-\bnabla \bcdot \left( \widetilde{ \widetilde\uu  \widetilde\uu} + \widetilde{\uu' \uu'} \right)
+\widetilde\ff,
\label{eq:fluct}
\end{align}	
where
$\LL(\UUm)$ is the Navier--Stokes operator linearised around the mean flow:
\be
\LL(\UUm) \widetilde\uu = (\UUm \bcdot \bnabla) \widetilde\uu + (\widetilde\uu \bcdot \bnabla)\UUm +\frac{1}{\rho} \bnabla \widetilde p - \nu \bnabla^2 \widetilde\uu.
\ee
Equation (\ref{eq:mean}) shows that,
due to the forcing from the 
 coherent and turbulent Reynolds stresses $\widetilde\uu \widetilde\uu+\uu'\uu'$,
  the mean flow $\UUm$ differs from the steady base flow $\UUbase$ which is a solution of the stationary Navier--Stokes equations $\NN(\UUbase) = \overline{\FF}$.

Focusing on a coherent time-harmonic forcing at frequency $\omega_1$,
\begin{align}	
\overline\FF=\zzz,
\quad
\widetilde\ff(\xx,t) = \widetilde\ff_{1}(\xx) e^{i \omega_1 t}+c.c.,
\quad \ff'=\zzz,
\end{align}	
where $c.c.$ stands for complex conjugate,
the coherent response is assumed to fluctuate at the forcing frequency and higher harmonics:
\begin{align}	
\widetilde\uu(\xx,t) 
= \sum_{n \neq 0} \widetilde\uu_{n}(\xx) e^{i n \omega_1 t}
= \sum_{n>0}      \widetilde\uu_{n}(\xx) e^{i n \omega_1 t}+c.c.
\quad (\widetilde\uu_{-n} = \widetilde\uu_{n}^*).
\label{eq:u_Fourier}
\end{align}
Introducing this Fourier decomposition into (\ref{eq:mean})-(\ref{eq:fluct})
yields an infinite system of equations for the mean flow $\UUm$ and each coherent fluctuation $\widetilde\uu_{n}$:
\begin{align}
\NN(\UUm) &= 
-\sum_{n \neq 0} \bnabla\bcdot  \widetilde\uu_{n}  \widetilde\uu_{-n} 
-\bnabla\bcdot \overline{\uu' \uu'},
\label{eq:mean_Fourier}
\\
i n \omega_1 \widetilde\uu_{n} + \LL(\UUm) \widetilde\uu_{n}
&= 
-\sum_{m \neq n,0} \bnabla\bcdot \widetilde\uu_{m} \widetilde\uu_{n-m}
-\bnabla\bcdot ( \widetilde{\uu'\uu'} )_n
+ \delta_{n1} \widetilde\ff_{1},
\label{eq:fluct_Fourier}
\end{align}	
where $( \widetilde{\uu'\uu'} )_n$ denotes the coherent component at frequency $n\omega_1$ of the turbulent Reynolds stresses,
and $\delta_{n1}$ is the Kronecker delta (equal to $1$ if $n=1$, and to $0$ otherwise).
For instance, the mean flow and first two harmonics are governed by 
\begin{align}
\NN(\UUm) &= 
-(\widetilde\ppsi_{1,-1}   
 +\widetilde\ppsi_{2,-2} \ldots) -\bnabla\bcdot \overline{\uu' \uu'},
\label{eq:mean_Fourier_ex}
\\
(i \omega_1 + \LL(\UUm)) \widetilde\uu_{1} &=
-(\widetilde\ppsi_{2,-1}
 +\widetilde\ppsi_{3,-2} \ldots)
-\bnabla\bcdot ( \widetilde{\uu'\uu'} )_1
+\widetilde \ff_{1},
\label{eq:fluct_Fourier_ex_1}
\\
(2 i \omega_1 + \LL(\UUm)) \widetilde\uu_{2} &= 
-(\widetilde\ppsi_{ 1 ,   1} \,\,\,\,
 +\widetilde\ppsi_{ 3 ,  -1} \ldots)
-\bnabla\bcdot ( \widetilde{\uu'\uu'} )_2,
\label{eq:fluct_Fourier_ex_2}
\end{align}	
where we have introduced the notations
\begin{align}
\widetilde\ppsi_{j,k} = \bnabla\bcdot \widetilde\uu_{j} \widetilde\uu_{k} + c.c.
\mbox{ if } j \neq k, 
\qquad
\widetilde\ppsi_{j,j} = \bnabla\bcdot \widetilde\uu_{j} \widetilde\uu_{j}, 
\end{align}
for the divergence of the coherent Reynolds stresses.
At this stage, these equations are exact.

The effect of the unknown turbulent Reynolds stresses $\uu'\uu'$ on coherent fluctuations in (\ref{eq:fluct_Fourier}) is accounted for with a turbulence model that relates $\widetilde{\uu' \uu'}$ to $\widetilde\uu$ via a turbulent viscosity $\nu_t$ (see details in \S~\ref{sec:Numerical}), such that (\ref{eq:fluct_Fourier_ex_1}) becomes
\begin{align}
(i \omega_1 + \LL(\UUm)) \widetilde\uu_{1} &=
-(\widetilde\ppsi_{2,-1}
 +\widetilde\ppsi_{3,-2} \ldots)
 +\widetilde \ff_{1},
\label{eq:fluct_Fourier_steady}
\end{align}	
where $\LL$ now contains the modified viscosity $\nu+\nu_t$.

We now make the central assumption that nonlinear forcing terms can be neglected (even though higher harmonics themselves may not be negligible), and that coherent fluctuations at $\omega_1$ can be predicted by the linear response:
\begin{align}
(i \omega_1 + \LL(\UUm)) \widetilde\uu_{1} &=
\widetilde \ff_{1}.
\label{eq:LNSE}
\end{align}
In the following, we will assess the ability of this simplified model to capture correctly amplification and saturation.

\subsection{Numerical method}
\label{sec:Numerical}

The two-dimensional linear response to time-harmonic forcing is calculated around the mean flow obtained from LES (\S~\ref{sec:Mean}).
The LNSE (\ref{eq:LNSE}) are recast in variational form and discretised 
in domain $I$ with the finite-element software \textit{FreeFem++} \cite{freefem}, using P2 and P1 Taylor–Hood elements for velocity and
pressure respectively (\cite{Boujo13, Boujo15a}). 
The two-dimensional mesh contains approximately 330 000 triangular elements, strongly clustered in the mixing layer.
See Appendix~A for a convergence study assessing the influence of the mesh size. 
Boundary conditions are as follows: $\widetilde\uu_1=\zzz$ at the inlet $\Gamma_{in}$ and on the walls $\Gamma_w$, 
stress-free condition $−\frac{1}{\rho}\widetilde p_1 \nn + (\nu+\nu_t) \widetilde \SS_1 \bcdot \nn = 0$ at the outlet $\Gamma_{out}$, and spatially uniform 
 vertical forcing 
$\widetilde\uu_1 = v' \eee_y$ at the cavity end $\Gamma_{f}$. 

The effect of turbulent fluctuations is taken into account with a classical eddy viscosity model. The coherent component of the turbulent Reynolds stresses $\widetilde{\uu'\uu'}$ is assumed to be proportional to the coherent strain rate 
$\widetilde{\SS} = \bnabla\widetilde\uu + \bnabla\widetilde\uu^T$ 
(Boussinesq approximation):
\be
\widetilde{\uu'\uu'} - \frac{2}{3} \widetilde{q} \II =  -2\nu_t \widetilde\SS,
\label{eq:boussinesq}
\ee
where $\widetilde{q}=\widetilde{\uu'\bcdot\uu'}/2$ is the kinetic energy and $\II$ the identity tensor.
In this study the turbulent viscosity $\nu_t$ is taken as space dependent, and calculated at each location according to
\be
\nu_t(\xx) = -\dfrac{\overline{\uu'\uu'} \bcol \overline\SS}{2\overline\SS \bcol \overline\SS },
\label{eq:turb_visc}
\ee
where $\bcol$ denotes the Frobenius inner product. 
(In other words, at each location $\nu_t$ can be seen as resulting from the least-square minimisation of the over-determined system of equations $\overline{\uu'\uu'} =  -2\nu_t \overline\SS$.)
The steady component of the turbulent Reynolds stresses $\overline{\uu'\uu'}$ and the mean strain rate $\overline\SS = \bnabla\overline\UU + \bnabla\overline\UU^T$ are evaluated from the LES.

As mentioned in Pope \cite{Pope00}, the above model is a two-fold simplification:
\textit{``First, there is the intrinsic assumption that (at each point and time) the Reynolds-stress anisotropy $\widetilde{\uu'\uu'} - \frac{2}{3} \widetilde{q} \II$ is determined by the mean velocity gradients.
Second, there is the specific assumption that the relationship (...)   is (\ref{eq:boussinesq}). This is, of course, directly analogous to the relation for the viscous stress in a Newtonian fluid.
(...)
In general, the turbulent viscosity hypothesis is incorrect. These general objections notwithstanding, there are important particular flows for which the hypothesis is more reasonable. In simple turbulent shear flows (e.g., the round jet, mixing layer, channel flow, and boundary layer) the turbulence characteristics and mean velocity gradients change relatively slowly (following the mean flow). As a consequence, the local mean velocity gradients characterise the history of the mean distortion to which the turbulence has been subjected; and the Reynolds-stress balance is dominated by local processes (...), the non-local transport processes being small in comparison. In these circumstances, then, it is more reasonable to hypothesise that there is a relationship between the Reynolds stresses and the local mean velocity gradients.''}

It should be noted that, in (\ref{eq:turb_visc}),
$\overline\SS$ is a direct output of the LES, whereas the calculation of $\overline{\uu'\uu'}$ requires further processing: indeed, the total statistics of the velocity field contain the steady component of both turbulent and coherent Reynolds stresses $\overline{\uu'\uu'} + \overline{\widetilde\uu \widetilde\uu}$.
This can be overlooked if the stresses $\overline{\widetilde\uu \widetilde\uu}$ are small compared to $\overline{\uu'\uu'}$ (\cite{Kitsios10,Viola14}), which is not the case here because substantial coherent fluctuations are produced by the harmonic forcing in the large-amplitude regime. Therefore, before computing the turbulent viscosity, we first remove the coherent contribution $\overline{\widetilde\uu_1 \widetilde\uu_{-1}}$  at the fundamental frequency, with $\widetilde\uu_{\pm1}$ obtained from  frequency analysis.
Some authors have used an alternative approach based on energetic structures obtained from proper orthogonal decomposition (\cite{Tammi2016}) or have proposed weighting the turbulent viscosity by a laminar-turbulent intermittency factor (\cite{Oberleithner14}).

\subsection{Unforced case: linear stability analysis and linear harmonic response}
\label{sec:ResultsUnforced}

\begin{figure}[] 
\centerline{
\includegraphics[height=12.7cm]{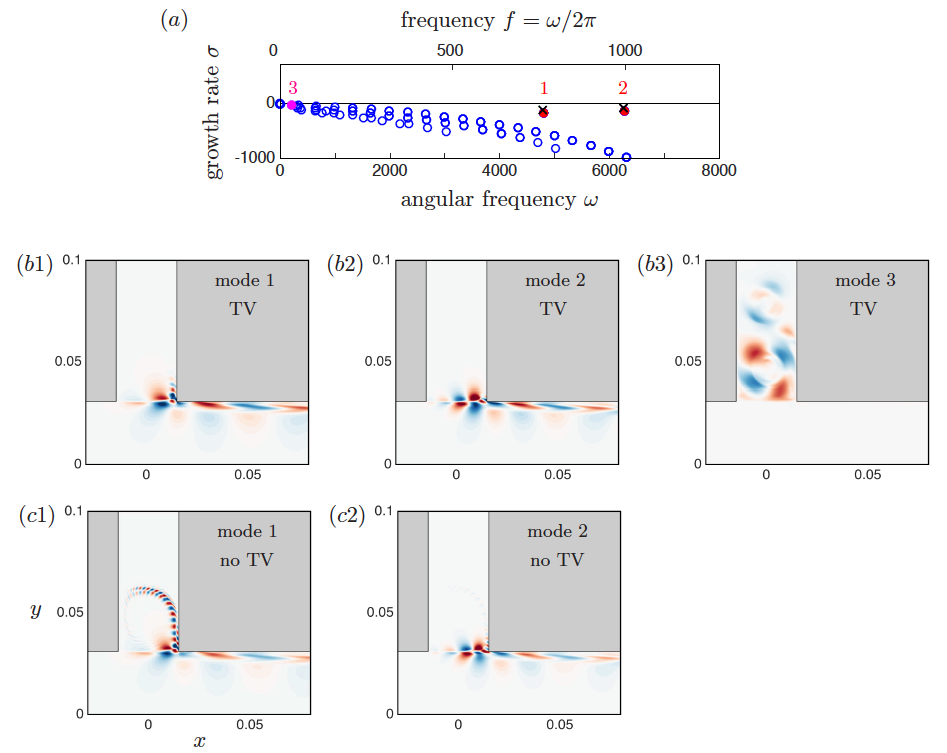}
}
\caption{
Global linear stability around the unforced mean flow ($v'=0$).
$(a)$~Eigenvalues lie on continuous branches, except marginally stable eigenvalues 1 and 2 which stand out at $\omega/2\pi=750$ and 1000~Hz.
These two eigenvalues are not substantially affected by discarding turbulent viscosity (crosses).
$(b,c)$~Eigenmodes (streamwise component, real part).
Modes 1 and 2  are located in the shear layer and the downstream boundary layer, computed either with turbulent viscosity (``TV'', panels $b1$, $b2$) or without (``no TV'', panels $c1$, $c2$).
$(b3)$ Other modes are located inside the cavity.
}
\label{fig:eigvals_eigmodes}
\end{figure}

We first focus on the unforced case, $v'=0$. Before moving to the harmonic response problem, we investigate linear stability by solving the eigenvalue problem 
\begin{align}
(\sigma + i \omega) \widetilde\uu 
+ \LL(\UUm) \widetilde\uu &= \zzz
\label{eq:linstab}
\end{align}
for infinitesimal perturbations 
$\widetilde\uu$ 
around the LES mean flow $\UUm$, with the  numerical method as described in \S~\ref{sec:Numerical}.
As shown in the spectrum in figure~\ref{fig:eigvals_eigmodes}$(a)$, all eigenmodes are stable (growth rate $\sigma \leq 0$). Most eigenvalues fall on continuous branches, and are more stable at larger frequencies. 
Two eigenmodes (denoted 1 and 2) stand out, however, at frequencies close to 750 and 1000~Hz, and are marginally stable (small growth rate compared to the angular frequency, $|\sigma| \ll \omega$). 
Marginal stability in mean flows has been observed in some laminar and turbulent flows, as well as counter-examples (\cite{Barkley06, Sipp2007, Turton15, Meliga2017JFM}).
Modes 1 and 2 are located in the shear layer and the downstream boundary layer, and exhibit approximately one and two wavelengths across the cavity, respectively, as shown in fig.~\ref{fig:eigvals_eigmodes}$(b1,b2)$. Other modes are located inside the cavity,  e.g. mode 3 in fig.~\ref{fig:eigvals_eigmodes}$(b3)$.
Discarding turbulent viscosity in the linear stability analysis (i.e. in $\LL$ in the eigenvalue problem (\ref{eq:linstab})) does not affect substantially  eigenvalues 1 and 2 (black crosses in panel $a$), and has a limited impact on the structure of  modes 1 and 2 in the shear layer (panels $c1$, $c2$).

Next, the linear response of the mean unforced  flow to harmonic forcing on $\Gamma_f$
is characterised in terms of kinetic energy in domain $\dom$ with the gain
\be 
G(\omega) = \dfrac{1}{v'} \left( 
\iint_\dom |\widetilde\uu_1|^2 \,\boldsymbol{\mathrm{d}}\xx 
\right)^{1/2}.
\ee
Figure~\ref{fig:freqresp}$(a)$ shows the gain obtained from the LNSE over a broad range of forcing frequencies (dashed line).
The forcing is amplified preferentially in the frequency range 600-1200~Hz, with clear peaks close to 750 and 1000~Hz that can be related to the marginally stable modes~1 and~2.

\subsection{Forced cases, saturation of the  linear harmonic response}
\label{sec:ResultsForced}

\begin{figure}[] 
\centerline{
\includegraphics[height=6cm]{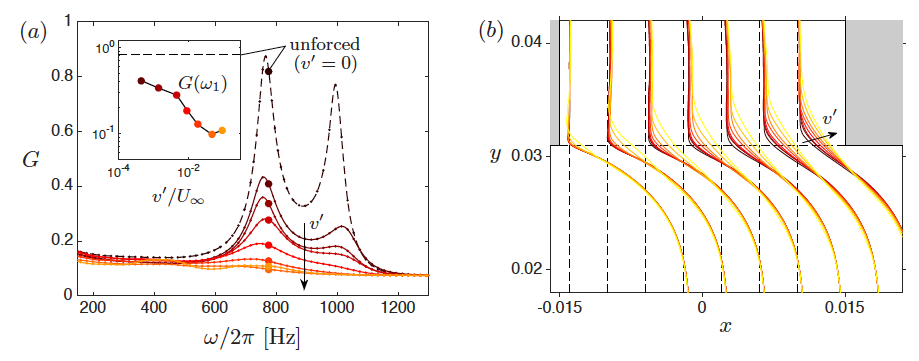}
}
\vspace{-0.2cm}
\caption{
$(a)$~Linear harmonic gain $G(\omega)$ of the turbulent mean flow forced at $\omega_1/2\pi=750$~Hz at different amplitudes $v'$. Inset: $G(\omega_1)$ 
in logarithmic scale.
$(b)$~Profiles of mean streamwise velocity $\Um$.
} 
\label{fig:freqresp}
\end{figure}

We now turn our attention to mean flows obtained from LES with harmonic forcing at $\omega_1/2\pi=750$~Hz, and recompute the linear response to harmonic forcing.
In these mean flows, the gain consistently decreases  with the forcing amplitude (solid lines in fig.~\ref{fig:freqresp}$a$).
 
In the following, we focus on the linear response to harmonic forcing at the frequency $\omega_1$ of the dominant peak.
At this specific frequency (inset), the gain decreases by approximately one order of magnitude from the unforced regime to the large-amplitude forcing regime $v' \gtrsim 2.5$~m/s.
It should be noted that this effect comes entirely from the mean flow, which varies with $v'$. Specifically, the mixing layer becomes thicker as $v'$ increases (fig.~\ref{fig:freqresp}$b$), suggesting that the weaker shear is the main cause for the reduced amplification.

\begin{figure}[] 
\centerline{
\includegraphics[height=7cm]{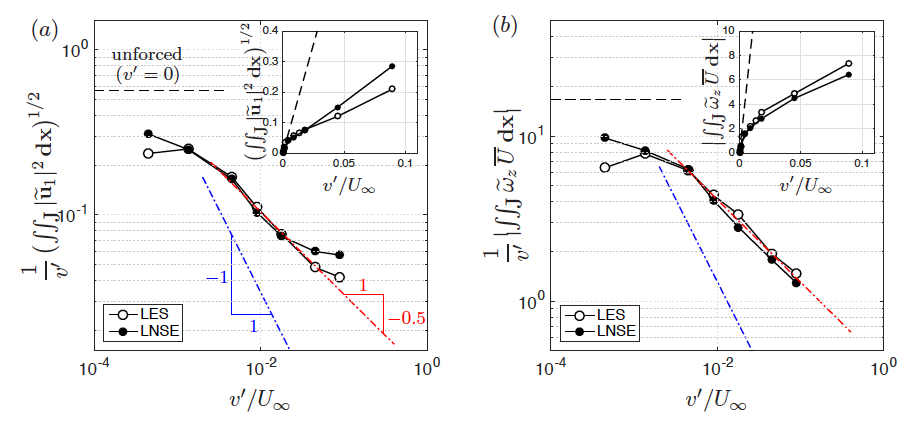}
}
\vspace{-0.2cm}
\caption{
Harmonic gain (logarithmic scale) at 
$\omega_1/2\pi=750$~Hz  vs. forcing amplitude, from LES (open symbols) and LNSE (filled symbols). 
The gain is calculated with the response measured in terms of 
$(a)$~kinetic energy (\ref{eq:gain_KE_subdom}) and  
$(b)$~dominant contribution of the vertical component of the unsteady Coriolis force (\ref{eq:G_coriolis}).
The linear gain for the unforced flow ($v'=0$) is shown as a horizontal dashed line.
Also shown are the slope -0.5 (red) that fits the data for $v'/\Ubulk\geq 0.45\%$, and the slope -1 (blue) that would be obtained for full saturation.
Insets: harmonic response (linear scale).
} 
\label{fig:freqresp2}
\end{figure}

Next, we take a closer look at the rectangular subdomain 
$\subdom=\{ (x,y) | -W/2 \leq x \leq W/2, 0\leq y \}$  spanning exactly the streamwise extension of the cavity.
Figure~\ref{fig:freqresp2}$(a)$ compares the harmonic gain restricted to region $\subdom$, obtained from LES and LNSE at $\omega_1$:
\be 
\dfrac{1}{v'} \left( 
\iint_\subdom |\widetilde\uu_1|^2 \,\boldsymbol{\mathrm{d}}\xx
\right)^{1/2}.
\label{eq:gain_KE_subdom}
\ee
The overall agreement is very good, with LNSE capturing well the decrease in gain observed in the LES.
The slight discrepancy at very low forcing amplitude ($v'=0.025$~m/s) can be ascribed to the small signal-to-noise ratio in the LES, which makes it difficult to  to measure the gain  accurately.
In the large-amplitude forcing regime, LNSE overestimates the coherent response, which points to non-negligible contributions from higher harmonics and/or to a deteriorating turbulent viscosity model.
The inset illustrates the saturation of the response itself (prior to normalisation by $v'$), with the slope quickly departing from the LNSE result obtained with the unforced flow (dashed line).

Figure~\ref{fig:freqresp2}$(b)$ shows an alternative measure of the gain, useful in an aeroacoustic context: we define
\be   
\dfrac{1}{v'}
 \left| \iint_\subdom \widetilde\omega_{z} \, \overline U \,\boldsymbol{\mathrm{d}}\xx \right|,
\label{eq:G_coriolis}
\ee
where $\widetilde\omega_{z} = \partial_x \widetilde v - \partial_y \widetilde u$ is the spanwise vorticity of the coherent response.
This measure gives insight into vortex sound production, since the Coriolis force $\totalvortvec \times \UU$ is related to the acoustic power $\mathcal{P}$ of a low-Mach number compact vorticity distribution, as expressed for instance by Howe's formula~\cite{Howe80}
\begin{align}
\mathcal{P} = -\iint \overline{\rho} (\totalvortvec \times \UU) \bcdot \uu_{ac} \,\boldsymbol{\mathrm{d}}\xx,
\end{align}
where $ \UU$ is the total unsteady velocity field,
$\totalvortvec$ the total vorticity field,
and $\uu_{ac}$ the acoustic (irrotational) component of the fluctuation.
In the present configuration, the above expression is well approximated by the contributions from the vertical component $v_{ac}$ of the acoustic fluctuations and from the vertical component $\totalvortscal_{z} \, U$ of the Coriolis force. 
In addition, the dominant contribution to the time-averaged power comes from
$\widetilde\omega_{z} \, \overline U$,
hence our choice for (\ref{eq:G_coriolis}).
(See \cite{Bauerheim17} for an in-depth analysis and discussion.)
Here again, the agreement between LES and LNSE results is very good, which suggests that the linearised approach can provide useful quantitative estimates of the produced acoustic power.

Both gains  (\ref{eq:gain_KE_subdom}) and (\ref{eq:G_coriolis}) decrease like $\sim 1/\sqrt{v'}$, as shown by the red line of slope -0.5. This is weaker than full saturation, which would yield $\sim 1/v'$ (blue line of slope -1), i.e. a response not increasing at all when the forcing amplitude increases. 
\cite{Graf10} and \cite{Nakiboglu12} reported a similar saturation slope of approximately $-0.6$ in other cavity flows (deep circular cavity $W/H=0.08$ and shallow axisymmetric cavity $W/H=1.48$, respectively).

\begin{figure}[] 
\centerline{
\includegraphics[height=12.7cm]{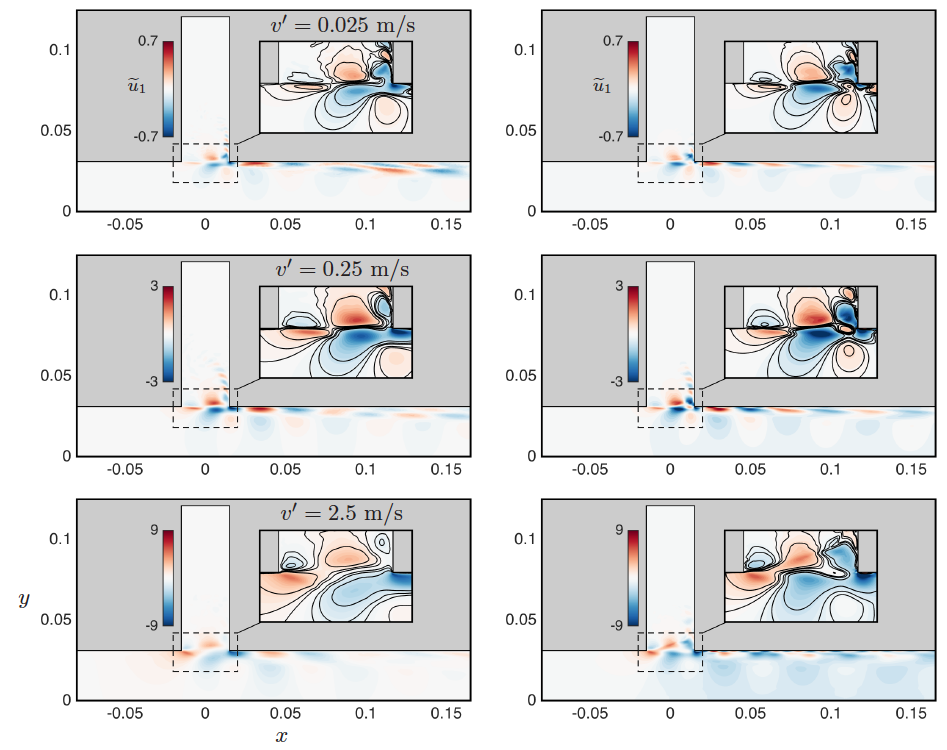}}
\vspace{-0.2cm}
\caption{
Streamwise component of $\widetilde\uu_1$ at $\omega_1$.
Left: LES spectral component.
Right: LNSE linear harmonic response around the mean flow.
Forcing amplitudes $v'=0.025$, 0.25 and 2.5~m/s.
} 
\label{fig:resp-u}
\end{figure}

The spatial structure of the harmonic response at $\omega_1$ is shown in figure~\ref{fig:resp-u}, obtained from the LNSE  calculation and from a spectral decomposition of the LES. 
The response is localised in the mixing layer, and in the downstream boundary layer. 
At low and medium forcing amplitudes, it has a distinct wave packet structure and experiences
a clear streamwise growth, with weak perturbations generated close to the upstream corner and amplified while convected by the mean flow to the downstream corner. This is typical of the linear response or instability of mixing layers.
At larger forcing amplitudes, the wave packet structure is less well organised and perturbations are convected with no substantial growth.
At this frequency, one wavelength  of the coherent response (i.e. one vortical structure) almost exactly fills the cavity width.
The shape and amplitude of the responses obtained from LNSE and LES are very similar, except in the vicinity of the downstream corner and for the largest forcing amplitudes.

\begin{figure}[] 
\centerline{
\includegraphics[height=10cm]{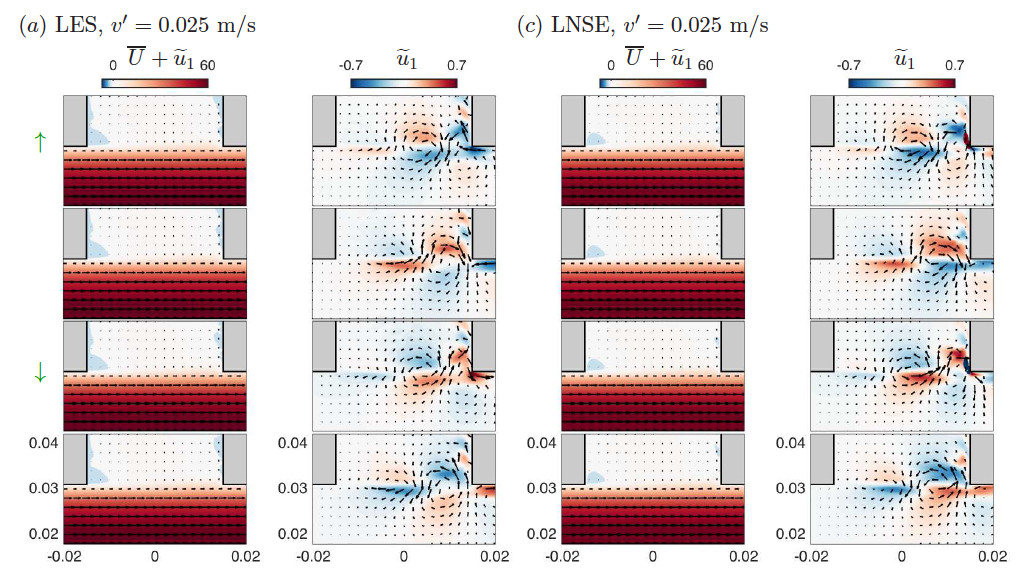}
}
\centerline{
\includegraphics[height=10cm]{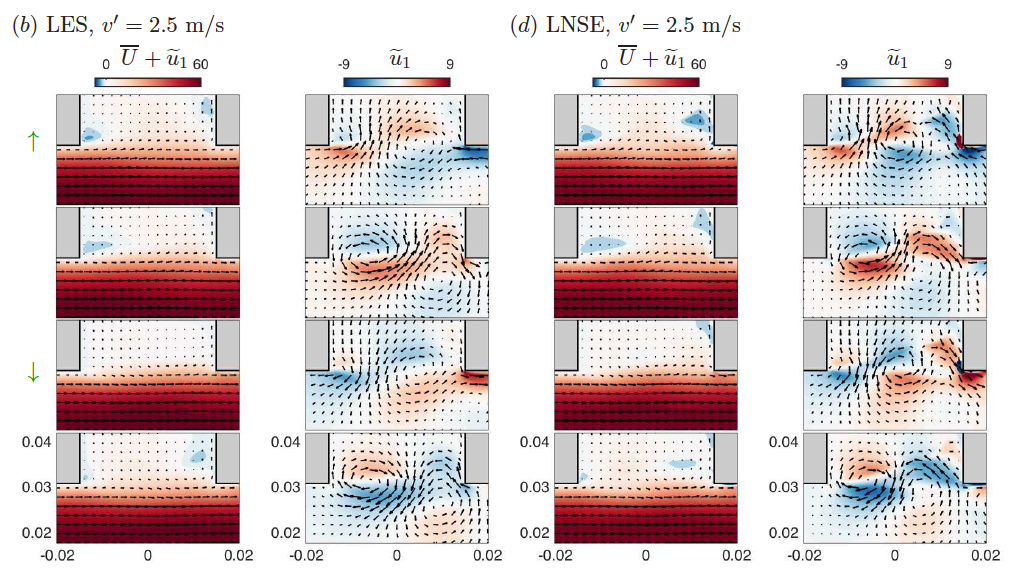}
}
\vspace{-0.2cm}
\caption{
Contours of streamwise velocity: 
phase-averaged velocity $\Um+\widetilde u_1$ and 
coherent velocity fluctuations $\widetilde u_1$.
(Overlaid are arrows of the phase-averaged and coherent fluctuating velocity fields $\UUm+\widetilde\uu_1$ and $\widetilde\uu_1$, respectively.)
$(a,b)$ LES, $(c,d)$ LNSE.
Forcing amplitudes: $(a,c)$ $v'=0.025$~m/s, $(b,d)$ $v'=2.5$~m/s.
Time instants in each panel, from top to bottom: 
$t$ (forcing directed upward $\uparrow$),
$t+T_1/4$, 
$t+T_1/2$ (forcing directed downward $\downarrow$) and 
$t+3T_1/4$.
} 
\label{fig:snaps-3}
\end{figure}

\begin{figure}[] 
\centerline{
\includegraphics[height=10cm]{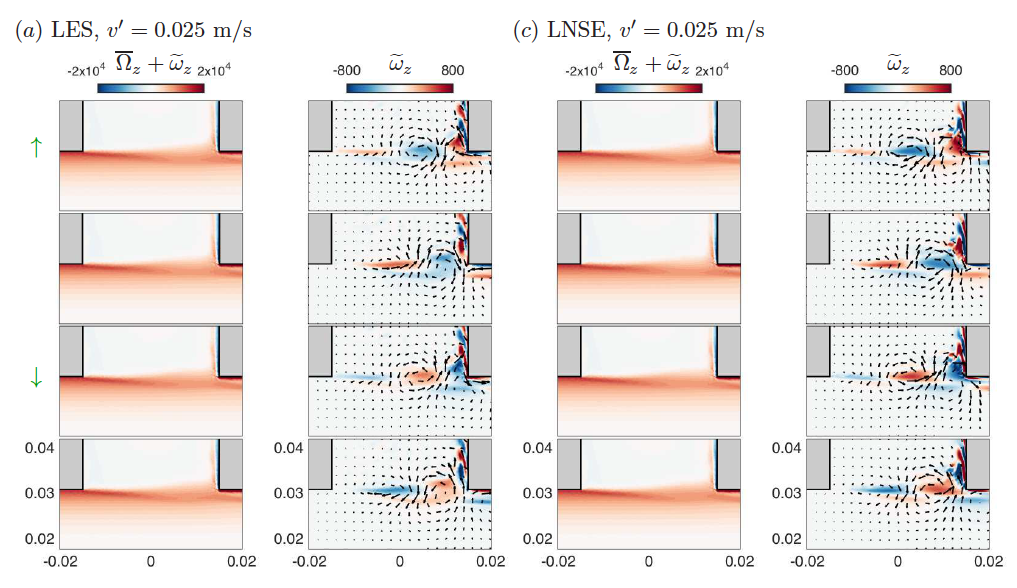}
}
\centerline{
\includegraphics[height=10cm]{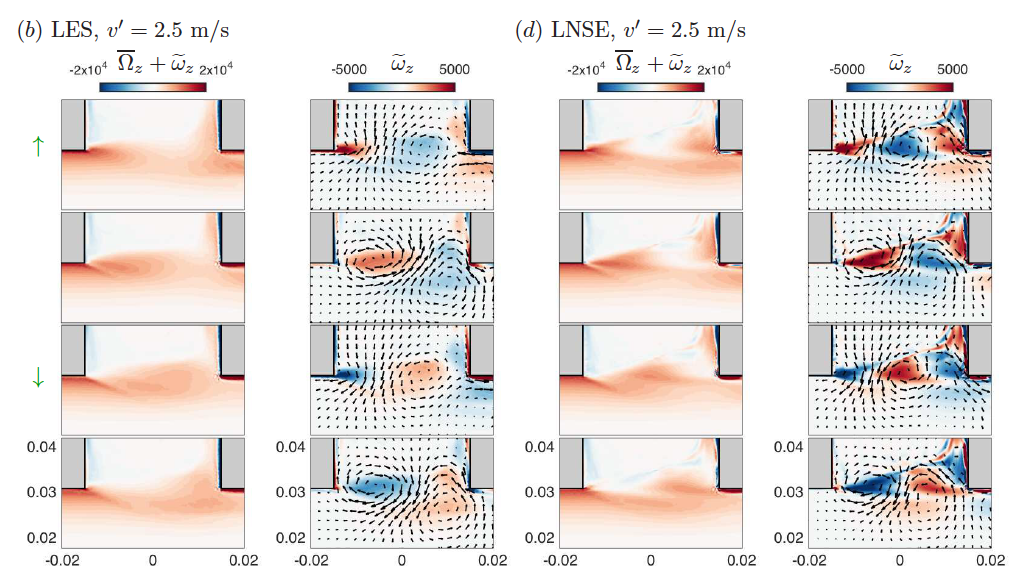}
}
\vspace{-0.2cm}
\caption{
Contours of spanwise vorticity: 
phase-averaged vorticity $\meanvort_z+\widetilde\omega_z$ and
coherent vorticity fluctuations $\widetilde\omega_z$.
(Overlaid are arrows of the coherent fluctuating velocity field $\widetilde\uu_1$.)
$(a,b)$ LES, $(c,d)$ LNSE.
Forcing amplitudes: $(a,c)$ $v'=0.025$~m/s, $(b,d)$ $v'=2.5$~m/s.
Time instants in each panel, from top to bottom: 
$t$ (forcing directed upward $\uparrow$),
$t+T_1/4$, 
$t+T_1/2$ (forcing directed downward $\downarrow$) and 
$t+3T_1/4$.
} 
\label{fig:snaps-4}
\end{figure}

Figure~\ref{fig:snaps-3} and \ref{fig:snaps-4} show snapshots of the response obtained from LES and LNSE at four time instants of an oscillation cycle separated by $T_1/4$, with $T_1=2\pi/\omega_1$ the forcing period.
At small forcing amplitude (panels $a$ and $c$), the gain is larger than at other forcing amplitudes but the resulting coherent response is small compared to the mean flow, and oscillations in the total flow are barely distinguishable. 
Snapshots of streamwise velocity and spanwise vorticity depict the formation and advection of two  main vortical structures of opposite vorticity in the first and second half-periods.
At larger forcing amplitude (panels $b$ and $d$), the amplification is substantially smaller but the coherent response is large enough that fluctuations in the total flow are visible.
One can clearly observe the role of the upstream corner in the formation of vortical structures, with $\omega_z>0$ when the forcing is directed upward (toward the cavity end) and, $T_1/2$ later,  $\omega_z<0$  when the forcing is directed downward (toward the main channel).

\begin{figure}[] 
\centerline{  
\includegraphics[height=5.7cm]{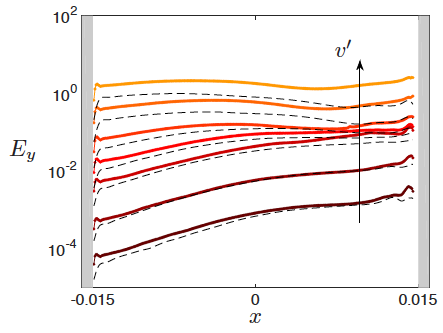}
}
\vspace{-0.2cm}
\caption{
Streamwise evolution of the energy density (\ref{eq:Ey}).
Thick solid lines: LNSE harmonic response;
Thin dashed lines: LES spectral component.
} 
\label{fig:resp-E}
\end{figure}

As observed in figures~\ref{fig:resp-u}-\ref{fig:snaps-3},
 the location of strongest response, where $\widetilde\uu_1$ is maximal, moves upstream as the forcing amplitude $|v'|$ increases.
This is further quantified in figure~\ref{fig:resp-E}, which represents the streamwise evolution of the coherent kinetic energy integrated vertically in $\subdom$ ($y>0$), i.e. the coherent energy density 
\be
E_y(x) = \int_{y>0} E(x,y) \,\mathrm{d}y 
= \frac{1}{2}\int_{y>0} |\widetilde u_1|^2+|\widetilde v_1|^2 \,\mathrm{d}y.
\label{eq:Ey}
\ee
At low forcing amplitudes, $E_y$ increases exponentially with $x$, consistent with the streamwise amplification mentioned earlier and with a linear amplification scenario, and reaches its maximum close to the downstream corner.
As the forcing amplitude increases, the region of exponential amplification becomes shorter and eventually vanishes; consequently, $E_y$ reaches its maximum already near the middle of the cavity, and eventually even in the first half.

\begin{figure}[] 
\centerline{
\includegraphics[height=12.7cm]{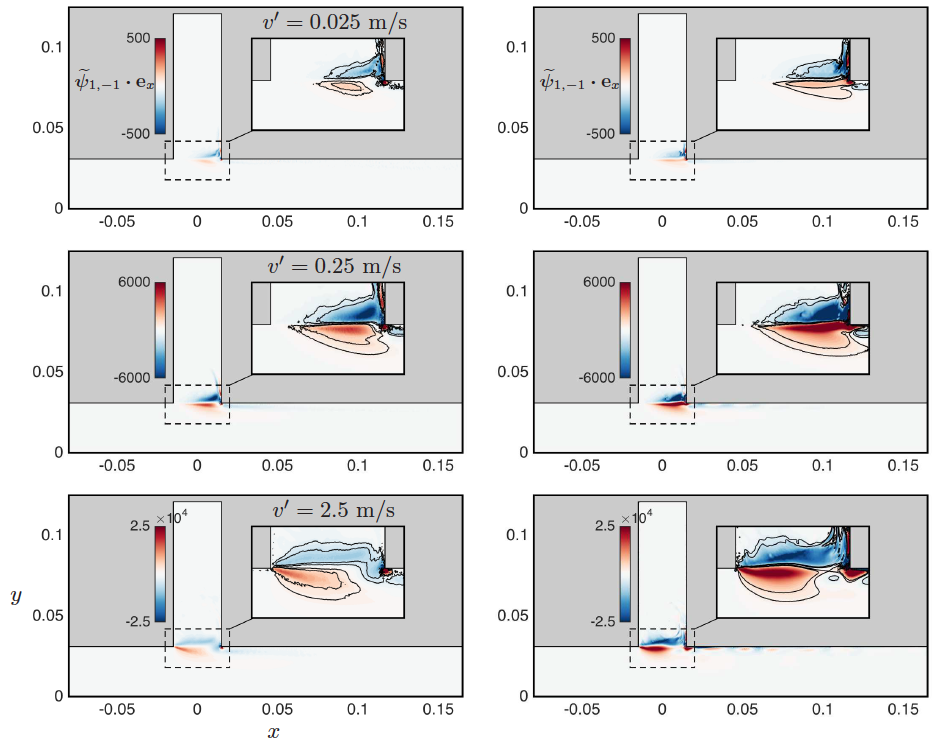}
}
\vspace{-0.2cm}
\caption{
Divergence of the coherent Reynolds stress, $\widetilde\ppsi_{1,-1}$ (vertical component), acting as a forcing term for the mean flow (see (\ref{eq:mean_Fourier_ex})).
Left: LES; right: LNSE.
Forcing at $\omega_1$, amplitudes $v'=0.025$, 0.25 and 2.5~m/s.
} 
\label{fig:RS}
\end{figure}

This upstream migration can be understood in terms of the mean flow distortion induced by the Reynolds stresses. As mentioned in \S~\ref{sec:LinResp}, the mean flow is forced by Reynolds stress divergence terms (see e.g. (\ref{eq:mean})).
As the forcing amplitude increases, coherent Reynolds stresses build up earlier upstream (figure~\ref{fig:RS}), leading to a thickening of the mean shear layer.
In turn, the coherent response building up around this increasingly diffused mean flow benefit from a reduced  potential for amplification, and saturate earlier upstream. 
This segregated yet coupled description, i.e. the interaction between (i)~the nonlinear mean flow forced by the coherent response and (ii)~the linear monochromatic coherent response around the mean flow, is the central ingredient of the simplified self-consistent model proposed to predict the saturation mechanism at play under harmonic forcing \cite{Mantic2016}.

\section{Validity of the monochromatic approximation}
\label{sec:ValidMonochrom}

\subsection{Amplitude of higher harmonics and corresponding forcing terms}

In the previous section we have considered the linear response at only one frequency, equal to the forcing frequency $\omega_1$, and we have focused on the external forcing $\widetilde\ff_1$, neglecting other forcing terms arising at $\omega_1$ from the interaction of higher-frequency coherent fluctuations, i.e. Reynolds stress divergence terms $\widetilde\ppsi_{j,1-j}$, $j\geq 2$, in (\ref{eq:fluct_Fourier_steady}).
The good agreement found between the response obtained in this LNSE framework and the flow computed with a fully nonlinear LES suggests that the nonlinear interaction of higher harmonics plays a negligible role 
on the response. (Note that this still requires the correct mean flow $\UUm$, which is crucially determined by $\widetilde\ppsi_{1,-1}$ and possibly influenced by higher-order terms $\widetilde\ppsi_{j,-j}$ as well. 
Here we use the fully nonlinear LES mean flow; a predictive method doing without direct nonlinear simulations would have to account for these Reynolds stresses carefully.)
In the following, we investigate this aspect further.

\begin{figure}[] 
\centerline{
\includegraphics[height=5.7cm]{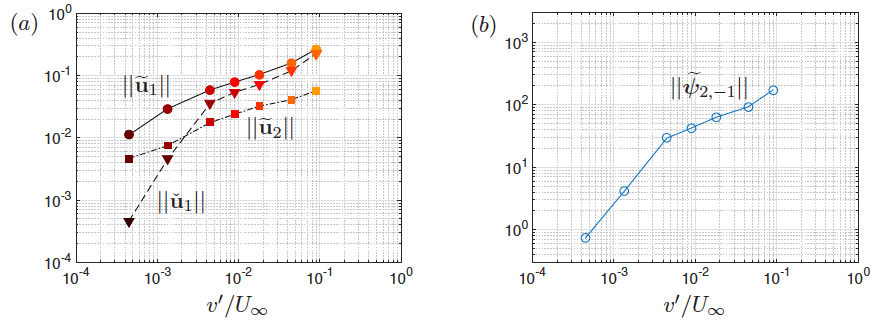}
}
\vspace{-0.2cm}
\caption{
$(a)$ 
Norm of the higher harmonic $\widetilde\uu_2$ at $2\omega_1$ (LES),
of the response $\widetilde\uu_1$ to the external forcing $\widetilde \ff_1$ at  $\omega_1$ (LES), 
and of the response $\check\uu_1$ to the first Reynolds stress divergence forcing term $\widetilde\ppsi_{2,-1}$ at  $\omega_1$ (LNSE).
$(b)$~Norm of $\widetilde\ppsi_{2,-1}$.
} 
\label{fig:norms_u1_u2_u1partial}
\end{figure}

In general, if higher harmonics are small, their interaction is necessarily small too.
We note that, in the present flow, the second harmonic $\widetilde\uu_2$ (extracted from the LES at $\omega_2=2\omega_1$) is smaller than $\widetilde\uu_1$ but far from negligible: the ratio $||\widetilde\uu_1||/||\widetilde\uu_2||$ is less than one order of magnitude (figure~\ref{fig:norms_u1_u2_u1partial}$a$).
In \cite{Turton15}, linear stability analysis around the mean flow in a laminar thermosolutal convection system reproduced well nonlinear characteristics for traveling waves whereas it failed for standing waves, which was explained by a negligible (resp. non-negligible) second harmonic in the former (resp. latter) case.

Higher harmonics, albeit not negligible, may still contribute only marginally to the harmonic response at $\omega_1$,  either 
(i) if their interaction as Reynolds stress divergence  $\widetilde\ppsi_{j,1-j}$ is small,
or 
(ii) if the response to these forcing terms is small.
Regarding condition (i), it would be natural to quantify rigorously what ``small $\widetilde\ppsi_{j,1-j}$'' means by comparing these 
forcing terms  to the external forcing $\widetilde\ff_1$; however, this not possible in the present configuration since the $\widetilde\ppsi_{j,1-j}$ are defined in the volume while $\widetilde\ff_1$ is applied at a boundary. 
Nonetheless, we report for the sake of completeness the norm of $\widetilde\ppsi_{2,-1}$, the first (and likely dominant) Reynolds stress divergence term, in figure~\ref{fig:norms_u1_u2_u1partial}$(b)$.
Regarding condition (ii), which may be verified irrespective of condition (i), comparing the response to each forcing term is straightforward; the question is therefore whether the response to the (possibly non-small) Reynolds stress forcing is small compared to the response $\widetilde\uu_1$  to the external forcing:
\be 
(i \omega_1 + \LL(\UUm))^{-1} \widetilde\ppsi_{j,1-j}
\ll 
(i \omega_1 + \LL(\UUm))^{-1} \widetilde\ff_1 \,?
\ee
Figure~\ref{fig:norms_u1_u2_u1partial}$(a)$ reports the norm of 
$\check\uu_1 = (i \omega_1 + \LL(\UUm)) \widetilde\ppsi_{2,-1}$. 
At the two lowest forcing amplitudes,  $\check\uu_1$ is indeed much smaller than $\widetilde\uu_1$ (in excess of 20 and 5 times, respectively). At larger amplitudes, however, both responses are of the same order of magnitude. At first glance, this seems at odds with the fact that $\widetilde\uu_1$ alone is sufficient to predict the overall coherent fluctuations at $\omega_1$. Interestingly, a closer look reveals that $\check\uu_1$ and $\widetilde\uu_1$  have different phases and therefore cannot interact constructively, meaning that the norm of the response is essentially unaffected by $\check\uu_1$.

\subsection{Optimal response: harnessing the potential for amplification}
\label{sec:optimal_response}

Further insight is gained by quantifying how efficiently the two forcing terms (external forcing, and forcing from the interaction of higher harmonics) are amplified. A direct comparison of the gains 
$||\widetilde\uu_1||/||\widetilde\ff_1||$ and
$||\check    \uu_1||/||\widetilde\ppsi_{2,-1}||$ gives little information, because the forcing terms are defined on a boundary and in the domain, respectively.
However, it is possible to assess whether each forcing efficiently exploits the  potential for amplification  available in the flow. 
In this context, it is natural 
to introduce the concept of  optimal gain, which corresponds to the largest possible linear amplification in the flow at a given frequency: instead of solving 
\begin{align}
(i \omega + \LL(\UUm))\uu &=\ff
\label{eq:LNSE2}
\end{align}
to compute the linear response $\uu$
to a \textit{given} harmonic forcing $\ff$ (defined either on a boundary or in the domain), the idea is to identify the \textit{optimal} forcing $\ff^{(opt)}$ which maximises the gain
\begin{align}	
G^{(opt)}(\omega) = \max_{\ff} \dfrac{||\uu||}{||\ff||} = \dfrac{||\uu^{(opt)}||}{||\ff^{(opt)}||} .
\label{eq:Gopt}
\end{align}
The optimal gain  is obtained by performing a singular value decomposition of the resolvent operator defined by 
\begin{align}
\uu = (i \omega + \LL(\UUm))^{-1} \ff = \RR(\omega)\ff,
\label{eq:resolv}
\end{align}
or, equivalently, by solving the eigenvalue problem $\RR^\dag \RR\ff = G^2\ff$ (where $\RR^\dag$ is the adjoint resolvent operator). 
If needed, the calculation actually yields more, namely an orthogonal set of optimal forcings $\ff^{(k)}$  and the corresponding set of optimal responses $\uu^{(k)}$ associated with optimal gains $G^{(k)}$ sorted in decreasing order:
\be 
G^{(opt)}=G^{(1)} \geq G^{(2)} \geq G^{(3)} \ldots
\ee
Examples of resolvent analyses around turbulent mean flows include \cite{Farrell1993, McKeon2013, Garnaud2013, Beneddine2016}.

\begin{figure}[] 
\centerline{  
\includegraphics[height=5.8cm]{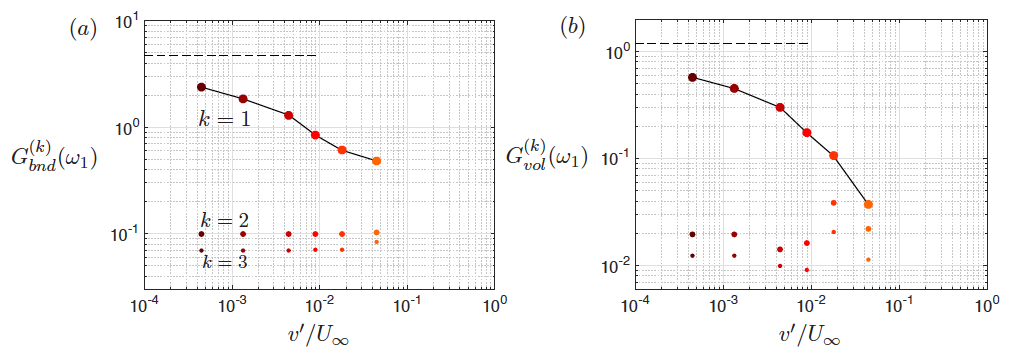}
}
\vspace{-0.2cm}
\caption{
First three optimal gains at $\omega_1$, for $(a)$~boundary forcing at the cavity end $\Gamma_f$, or $(b)$~volume forcing in the domain $\dom$.
} 
\label{fig:optfreqresp}
\end{figure}

Figure \ref{fig:optfreqresp} shows the first three optimal gains for harmonic forcing applied at $\omega_1$ at the cavity end or in the domain. 
At small forcing amplitudes, the first optimal gain is more than one order of magnitude larger than the  following optimal gains, which  has been observed in other flows  (\cite{Dergham13, Boujo15a, Beneddine2016}).
At larger amplitudes, this clear separation persists for boundary forcing, while for volume forcing the first optimal forcing becomes increasingly less amplified and is eventually comparable to the following optimal forcings.

\begin{figure}[] 
\centerline{
\includegraphics[height=12.7cm]{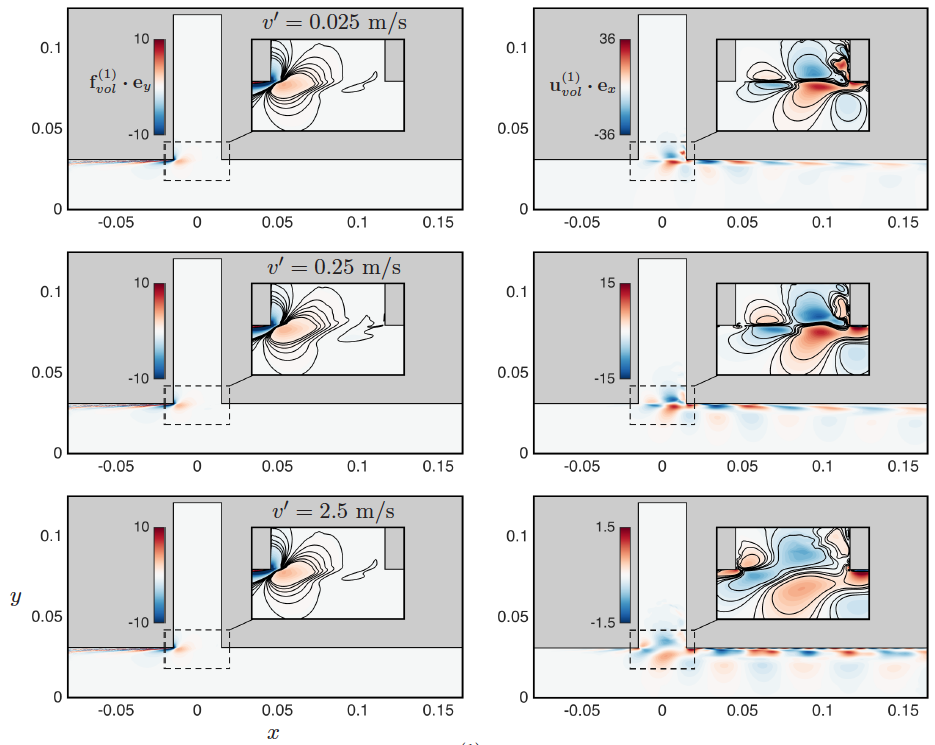} 
}
\vspace{-0.2cm}
\caption{
Left: 
optimal volume forcing $\ff_{vol}^{(1)}$ at $\omega_1$ (unit norm; vertical component);
Right: 
corresponding optimal response $\uu_{vol}^{(1)}$ (norm $G_{vol}^{(1)}$; streamwise component).
Mean flow at forcing amplitudes $v'=0.025$, 0.25 and 2.5~m/s.
} 
\label{fig:optresp}
\end{figure}

In both cases, the first optimal gain decreases with forcing amplitude, indicating that the dominant amplification mechanism weakens as the mean flow is modified. 
This is confirmed by the optimal volume forcing and
optimal response shown in figure~\ref{fig:optresp}: they clearly identify  the mixing layer as the main amplification region, and shear as the main amplification mechanism.

\begin{figure}[] 
\centerline{
\includegraphics[height=5.5cm]{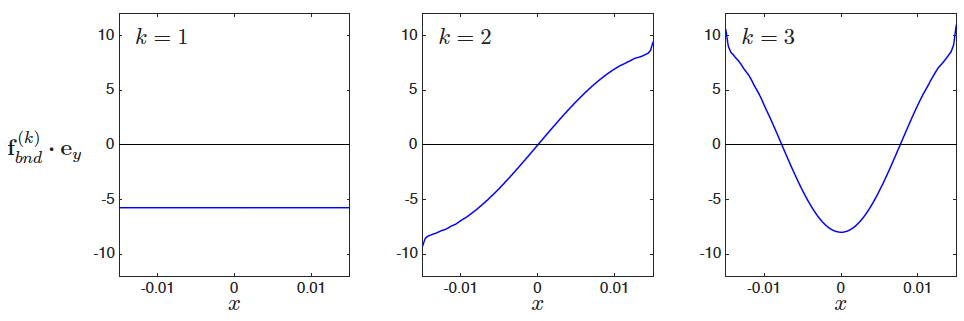}
}
\vspace{-0.2cm}
\caption{
First three optimal boundary forcings at the cavity end $\Gamma_f$ at $\omega_1$ (real part of vertical component of of unit-norm $\ff_{bnd}^{(k)}$).
Mean flow forced at amplitude $v'=0.075$~m/s.
} 
\label{fig:optforcfloor}
\end{figure}

The first three optimal boundary forcings at the cavity end $\Gamma_f$ are shown in figure~\ref{fig:optforcfloor} for $v'=0.075$~m/s (they are essentially independent of $v'$). 
They exhibit an increasing number of spatial oscillations over the cavity width. 
Interestingly, the first optimal is uniform, meaning that the external forcing $\widetilde\ff_1$ considered in this study is actually optimal.
The response to the optimal boundary forcing has  therefore the same structure as the response to $\widetilde\ff_1$ shown in figure~\ref{fig:resp-u}. 
One can observe that the optimal volume response and optimal boundary response have very similar structures at lower and intermediate forcing amplitudes, i.e. when the first optimal gain is much larger than the following optimal gains.

\begin{figure}[] 
\centerline{  
 \includegraphics[height=6.cm]{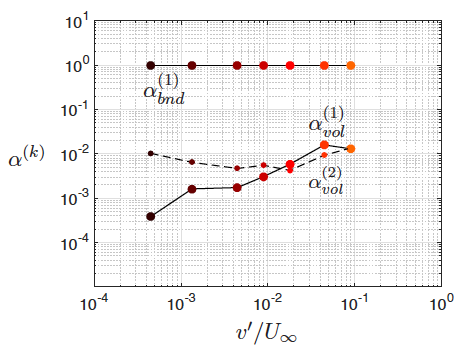}
}
\vspace{-0.2cm}
\caption{
Projection coefficients $\alpha^{(k)}$ on the $k$-th optimal volume forcing $\ff^{(k)}$ at $\omega_1$.
Coefficients $\alpha_{bnd}$ and $\alpha_{vol}$ correspond respectively to the external forcing $\widetilde\ff_1$ applied at the cavity end, and to  the volume Reynolds stress forcing $\widetilde\ppsi_{2,-1}$ resulting from higher-harmonic interactions. 
} 
\label{fig:scalprod_f_fopt}
\end{figure}

One might  wonder whether the volume forcing term from the Reynolds stress divergence $\widetilde\ppsi_{2,-1}$ is  close to the optimal volume forcing.
For a quantitative answer, let us decompose any forcing $\ff$ (at a boundary or in the domain) using the optimal forcings ($\ff_{bnd}^{(k)}$ or $\ff_{vol}^{(k)}$):
\begin{align}	
\ff = \sum_{k\geq 1}  \alpha^{(k)}  \ff^{(k)}.
\end{align}
Since the optimal forcings are orthogonal, the coefficients $\alpha_{bnd}^{(k)}$ and $\alpha_{vol}^{(k)}$ are easily expressed  in terms of a projection of the  considered forcing onto the optimal forcings:
\begin{align}	
\alpha^{(k)} = \dfrac{ \ps{\ff}{\ff^{(k)}} }{||\ff^{(k)}||^2}.
\end{align}
With this notation, the projection coefficients for the forcing $\widetilde \ff_1$ at the cavity end are therefore $\alpha_{bnd}^{(1)}=1$, and $\alpha_{bnd}^{(k)}=0$ for $k\geq 1$.
By contrast, the  projection coefficients $\alpha_{vol}^{(1)}$ and $\alpha_{vol}^{(2)}$ for the forcing arising from higher-harmonic interactions are of the order of  $10^{-3}$--$10^{-2}$ (figure~\ref{fig:scalprod_f_fopt}), meaning that $\widetilde\ppsi_{2,-1}$ projects very poorly on the first two optimal volume forcings.
In other words, the external forcing $\widetilde \ff_1$  harnesses \textit{all} the amplification available from the cavity end, while $\widetilde\ppsi_{2,-1}$ only benefits from 0.1--1$\%$ of the amplification available in the domain.

\begin{figure}[] 
\centerline{
\includegraphics[height=12.7cm]{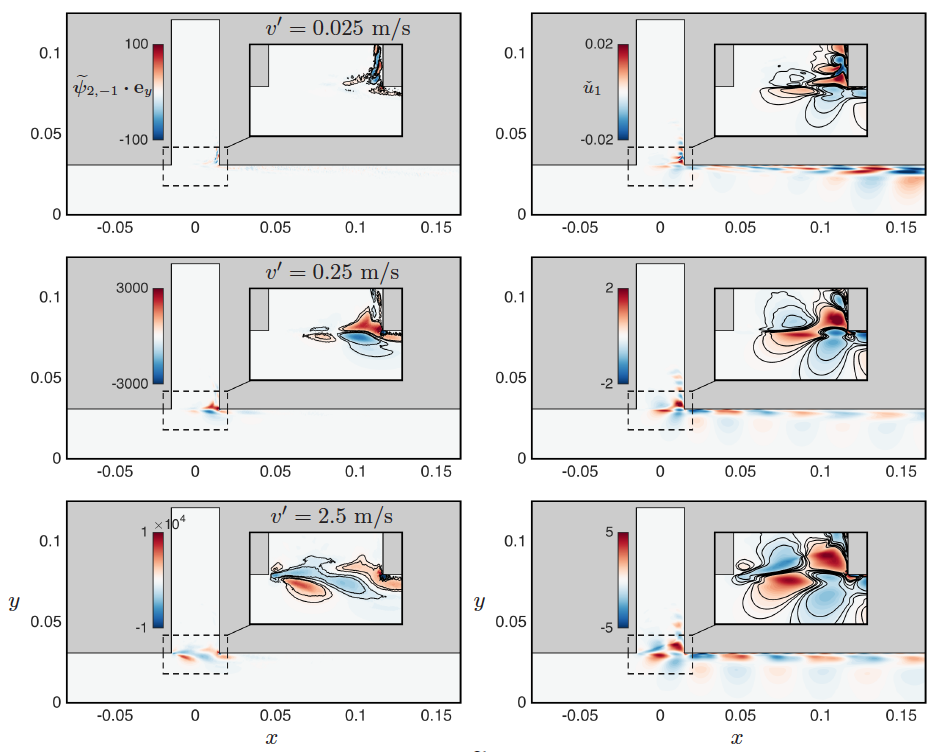}
}
\vspace{-0.2cm}
\caption{
Left: Reynolds stress divergence $\widetilde\ppsi_{2,-1}$  forcing the flow at $\omega_1$, from LES.
Right: linear response $\check \uu_{1}$ to $\widetilde\ppsi_{2,-1}$, from LNSE.
External forcing at $\omega_1$, amplitudes $v'=0.025$, 0.25 and 2.5~m/s.
} 
\label{fig:u2gradu1_and_resp}
\end{figure}

One reason for the poor amplification of $\widetilde\ppsi_{2,-1}$ is understood by inspecting its
spatial structure in figure~\ref{fig:u2gradu1_and_resp} and that of $\ff_{vol}^{(1)}$ in figure~\ref{fig:optresp}.
At small and intermediate forcing amplitudes,  
Reynolds stresses are concentrated in the downstream region of the cavity, while the optimal forcing is localised around the upstream corner.
At larger forcing amplitudes, the spatial overlap is better but the projection is still small because the structures remain essentially orthogonal: for instance, while $\ff_{vol}^{(1)}\bcdot\eee_y$ is uniform over the whole height of the shear layer, $\widetilde \ppsi_{2,-1}\bcdot\eee_y$ changes sign.

\section{Sensitivity analysis}
\label{sec:Sensitivity}

Using the linear response to harmonic forcing around the mean flow, we have gained understanding about amplification and saturation in the turbulent mixing layer over a deep cavity.
It is now natural to investigate flow control in order to reduce or increase the acoustic level. 
Considering the acoustic forcing $\widetilde \ff_1$ (frequency and spatial shape) as given, one possible strategy is to modify the mean flow $\UUm$ (e.g. using wall actuation or a passive control device), which in turn will modify the linear response $\widetilde \uu_1$.
How much the gain $G$ is affected by a \textit{given} flow modification can be found by recomputing the response.
However, a systematic study aiming at finding the most sensitive regions would imply a large computational cost.
A more efficient method consists in predicting the effect of \textit{any} small-amplitude flow modification using adjoint-based sensitivity.

\subsection{Adjoint-based sensitivity: background}

Sensitivity analysis was introduced in the context of hydrodynamic linear stability by \cite{Hill92AIAA}, and later used in various parallel (\cite{Bottaro03}), non-parallel two-dimensional (\cite{Giannetti07, Marquet08cyl, Meliga10}) and three-dimensional flows (\cite{Fani2012}) and in thermoacoustic systems (\cite{Magri13}) to compute the gradient 
\begin{align}
\bnabla_{*} \lambda
=
\dfrac{\mathrm{d}\lambda}{\mathrm{d}*}
\label{eq:grad}
\end{align}
of an eigenvalue 
$\lambda=\sigma+i\omega$ 
with respect to a variety of  modifications $(*)$:
(i)~steady flow modification,
(ii)~steady volume/boundary control,
(iii)~hypothetical localised ``velocity-to-force'' feedback (``structural sensitivity'').
(See  \cite{Chomaz05} for more details about structural sensitivity, and \cite{Luchini14} for a broad review about adjoint equations.)
The gradient (\ref{eq:grad}) is a useful information since it immediately predicts, via a simple scalar product, the first-order variation $\lambda\rightarrow\lambda+\delta\lambda$ induced by any small-amplitude modification: for instance, a flow modification $\UU \rightarrow \UU+\bdelta\UU$ induces an eigenvalue variation
$$
\delta\lambda 
= \ps{\bnabla_{\UU} \lambda}{\bdelta\UU}
= \ps{\dfrac{\mathrm{d}\lambda}{\mathrm{d}\UU}}{\bdelta\UU}.
$$

For linearly stable flows, \cite{Brandt11} extended sensitivity analysis to the linear response to harmonic \textit{volume} forcing.
They found that the sensitivity of the (squared) harmonic gain $G^2$ with respect to a modification of the flow $\UU$ (case (i) above) is given by 
\begin{align}
\bnabla_{\UU} G_{vol}^2
= 
2 G_{vol}^2 \mbox{Re} \left\{ 
-\bnabla\uu^H     \bcdot \ff_{vol} 
+\bnabla\ff_{vol} \bcdot \uu^*
\right\},
\label{eq:sensit_volume_forcing_flow_modif}
\end{align}
where $\uu$ is the response to the  volume forcing $\ff_{vol}$, and $(\cdot)^H$ denotes Hermitian transpose (conjugate transpose).
\cite{Boujo15a} considered \textit{boundary} forcing $\ff_{bnd}$, in which case the sensitivity of the harmonic gain to flow modification is
\begin{align}
\bnabla_{\UU} G_{bnd}^2
= 
2 \mbox{Re} \left\{ 
-\bnabla\uu^H \bcdot \uu^\dag 
+\bnabla\uu^\dag   \bcdot \uu^*
\right\},
\label{eq:sensit_boundary_forcing_flow_modif}
\end{align}
where $\uu$ is the response to the  boundary forcing $\ff_{bnd}$, and
the adjoint perturbation $\uu^\dag$ is a solution of the adjoint resolvent problem with $\uu$ as volume forcing (see Appendix B for details).
From the knowledge of (\ref{eq:sensit_volume_forcing_flow_modif}) or (\ref{eq:sensit_boundary_forcing_flow_modif}), one can proceed to compute the sensitivity to control (case (ii) above).

Recently, \cite{Qadri17} proposed an equivalent to structural sensitivity (case (iii) above) for the  amplification of harmonic \textit{volume} forcing.
Rearranging their expression, the overall gain variation  for a unit feedback localised in $\xx=\xx_0$ can be recast as 
\begin{align}
%
\delta (G_{vol}^2) 
= -2 G_{vol}^2 \mbox{Re} \left\{ \ff_{vol}(\xx_0) \bcdot  \uu(\xx_0) \right\}.
\label{eq:struct_sensit_volume_forcing}
\end{align}
When considering \textit{boundary} forcing, an additional intermediate step is necessary: in this case, the gain variation for a localised unit feedback is 
\begin{align}
%
\delta (G_{bnd}^2) 
= -2 \mbox{Re} \left\{ \uu^\dag(\xx_0) \bcdot  \uu(\xx_0) \right\},
\label{eq:struct_sensit_boundary_forcing}
\end{align}
where, again, $\uu^\dag$ is a solution of the adjoint resolvent problem with $\uu$ as volume forcing (see details in Appendix~B).
A simple way to analyse the sensitivity of the harmonic gain with respect to localised feedback is to look at the space-dependent product
\be 
||\ff_{vol}(\xx_0)|| \, ||\uu(\xx_0)|| 
\qquad \mbox{or} \qquad
||\uu^\dag(\xx_0)|| \, ||\uu(\xx_0)||, 
\ee
which is analogous to the structural sensitivity of an eigenvalue (product of the direct and adjoint modes).

\begin{figure}[] 
\centerline{
\includegraphics[height=12.7cm]{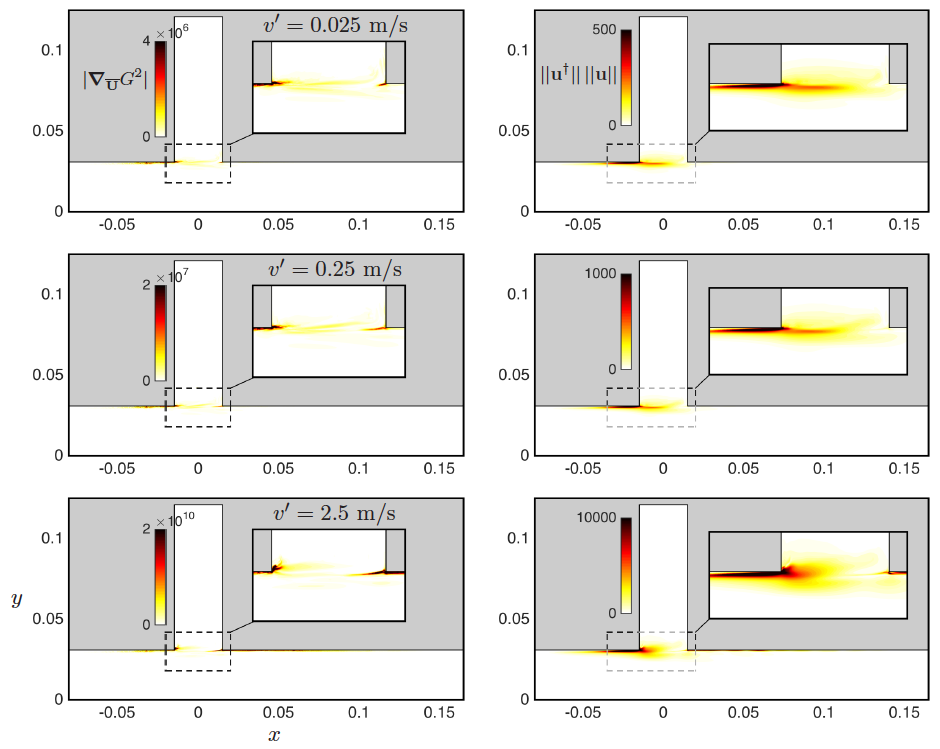}
}
\vspace{-0.2cm}
\caption{
Sensitivity of the linear optimal gain $G_{opt}^2(\omega_1)$, for harmonic forcing at the cavity end $\Gamma_f$.
$(a)$~Magnitude of the sensitivity with respect to mean flow modification (\ref{eq:sensit_boundary_forcing_flow_modif}).
$(b)$~Structural sensitivity (\ref{eq:struct_sensit_boundary_forcing}), i.e. sensitivity with respect to a localised feedback.
Forcing amplitudes $v'=0.025$, 0.25 and 2.5~m/s.
} 
\label{fig:wavemaker_and_sensit_BFmod_mag}
\end{figure}

\subsection{Sensitivity of the optimal gain}

In the following we investigate the sensitivity of the optimal harmonic gain for boundary forcing at the cavity end $\Gamma_f$, at frequency $\omega_1$. 
The sensitivity to mean flow modification (\ref{eq:sensit_boundary_forcing_flow_modif})
and structural sensitivity (\ref{eq:struct_sensit_boundary_forcing}) are shown in figure~\ref{fig:wavemaker_and_sensit_BFmod_mag}.
Both reach their largest magnitude in a localised region near the upstream cavity corner (and, to a lesser extent, in the boundary layer upstream of the cavity as well as in the mixing layer;
a second region of large sensitivity appears at  the downstream corner as the forcing amplitude $v'$ increases).
Therefore, the linear amplification between optimal boundary forcing  and optimal response is the most sensitive to modifications near the upstream corner.

This is consistent with the fact that most of the amplification is driven by the mean shear:
acoustic forcing induces perturbations near the upstream corner, which are then amplified in the  mixing layer.
Any control aiming at modifying the amplification should therefore target the upstream corner.
This also explains why saturation is more marked once fluctuations (Reynolds stresses) have moved upstream, in the region where the mean flow is more sensitive to their effect.

We recall from \S~\ref{sec:optimal_response} that 
(i)~the responses to optimal boundary and volume forcing are very similar (in terms of spatial structure), and that 
(ii)~the first optimal gain is almost always much larger that the following optimal gains.
These two elements point to a robust amplification mechanism in the mixing layer, fairly insensitive to the exact shape and location of the forcing.
In addition, we note that the adjoint perturbation $\uu^\dag$ in the sensitivities (\ref{eq:sensit_boundary_forcing_flow_modif}) and  (\ref{eq:struct_sensit_boundary_forcing}) looks similar to the optimal response $\ff^{(opt)}$, which leads to large sensitivities in the exact same regions for volume forcing (not shown) and boundary forcing (fig.~\ref{fig:wavemaker_and_sensit_BFmod_mag}).

\begin{figure}[] 
\centerline{   
\includegraphics[height=4.5cm]{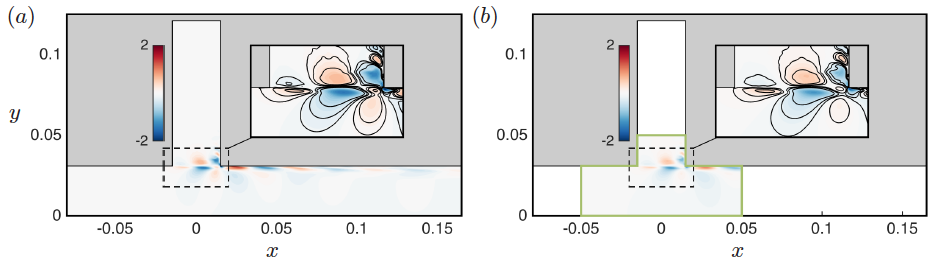}
}   
\vspace{-0.2cm}
\caption{
Effect of the size of the LNSE domain used to compute the linear response (frequency $\omega_1$, 
forcing amplitude $v'=0.075$~m/s).
$(a)$ Reference domain \dom.
$(b)$ Smaller domain D$^{xy}_3$.
} 
\label{fig:comp_resp_smaller_dom}
\end{figure}

Finally, we observe that the linear gain does not vary substantially when reducing the size of the computational domain, as long as the boundaries are not too close to the region of large  structural sensitivity (see details in Appendix~C). 
The spatial structure of the response is unaffected too (figure~\ref{fig:comp_resp_smaller_dom}). 
This is in agreement with \cite{Giannetti07}, who found a similar behaviour for the leading eigenvalue and eigenmode of the flow past a circular cylinder at $\Rey=50$:
they hypothesised that ``\textit{the characteristics of the global mode are dictated mainly by the conditions existing in the region where values of}'' structural sensitivity ``\textit{substantially different from zero are attained}'', i.e. where the the global mode and associated adjoint mode overlap.
We  formulate the same hypothesis in the case of  harmonic amplification: the value of the linear gain and the spatial structure of the response are dictated mainly by the flow in the region where  structural sensitivity is not small, i.e. where the the response $\uu$ 
and the volume forcing $\ff$ (or the adjoint perturbation $\uu^\dag$ associated with boundary forcing) overlap.

\section{Conclusion}
\label{sec:conclu}

We consider the turbulent flow over a deep cavity and compute the linear response to a uniform harmonic forcing applied at the cavity end. 
This forcing mimics a plane acoustic wave corresponding to the dominant  acoustic resonance mode (quarter-wave mode) at frequency $\omega_1$.
Calculations are carried out in the framework of the incompressible Linearised Navier--Stokes Equations (LNSE) with an eddy-viscosity turbulence model, and using as linearisation point the mean flow obtained from nonlinear Large-Eddy Simulations (LES) with a similar harmonic forcing at several amplitudes spanning more than two orders of magnitude. 
The influence of higher harmonics on the mean flow is automatically accounted for via LES, while their influence on coherent oscillations at $\omega_1$ is neglected.
The aim of the work is to assess the ability of this LNSE-based procedure to yield accurate results in terms of spatial structure and response amplitude, and to capture the saturation mechanism that, ultimately, would set the limit-cycle oscillation amplitude in a self-excited aeroacoustic resonance.

We find that the response amplitude is well predicted, both with a hydrodynamic measure (kinetic energy) and with an acoustic measure (Coriolis force involved in acoustic power generation). 
Vortical structures in the shear layer are in good agreement too, except at very large forcing amplitudes.
The gain (amplification of the forcing) is largest in the unforced case and decreases with forcing amplitude. 
This is consistent with the following saturation scenario: as the amplitude of oscillations grows, their nonlinear interaction (in the form of Reynolds stresses) modifies the mean flow, and shear-driven amplification in the thickened shear layer is reduced.
We observe that this good agreement is possible even though higher harmonics, neglected in the LNSE, are not small. This suggests that the mean flow contains all important nonlinearities.

We also note with a resolvent analysis that the optimal boundary forcing (t.e. the forcing which undergoes the largest possible amplification) at the cavity end is uniform, i.e. the forcing we prescribe in our study has precisely the optimal shape and benefits from the entire potential for amplification available at the cavity end. 
By contrast, the nonlinear interaction of the first and second harmonics which forces the flow at $\omega_1$ is projected poorly on the optimal volume forcing and does not take advantage efficiently of the potential for amplification available in the domain.

Finally, sensitivity analysis identifies  the upstream boundary layer and upstream cavity corner as regions where both localised feedback and mean flow modification have the largest effect on harmonic amplification, an information that can contribute to a systematic and computationally inexpensive control design.

A future goal is to extend the method to the prediction of limit-cycle amplitudes in self-excited aeroacoustic resonances. 
This should be accomplished in a stand-alone fashion, i.e. without relying on expensive nonlinear simulations such as LES to compute the mean flow. 
Extending semi-linear self-consistent models (\cite{Mantic2014, Mantic2016}) to turbulent flows would, if technically possible, constitute a promising method.

\section*{Appendix A. Convergence study on mesh size}

The influence of the mesh size on the LNSE results is analysed with a convergence study involving four meshes, M$_1$ to M$_4$. 
All meshes share the same structure: coarser at the inlet and outlet, and gradually finer toward the shear layer.
Mesh M$_1$ contains $N_{SL}=315$ vertices across the shear layer $\{ -W/2 \leq x \leq W/2, y=D/2 \}$, resulting in approximately $N_e$=120 000 triangular elements in the whole domain $\dom$.
Meshes M$_2$ to M$_4$ are obtained by applying to M$_1$ a uniform refinement of factor 1.33, 1.67 and 2, resulting in approximately $N_e$=204~000, 331~000 and 458~000 elements, respectively (table~\ref{tab:meshes}).

\begin{table}
  \begin{center}
\def~{\hphantom{0}}
  \begin{tabular}{ c | cccc }
          & M$_1$   & M$_2$   & \textbf{M$_3$}   & M$_4$    \\ \hline 
$N_{SL}$  &     315 &     420 &     \textbf{525} &     630 \\ 
$N_{v}$   &  60 743 & 103 045 & \textbf{166 464} & 230 240 \\
$N_{e}$   & 120 214 & 204 388 & \textbf{330 794} & 457 921 \\
  \end{tabular}
  \caption{Meshes used for the convergence study: 
number of vertices across the shear layer,
total number of vertices and total number of triangular elements.}
  \label{tab:meshes}
  \end{center}
\end{table}

\begin{figure}[] 
\centerline{
\includegraphics[height=6cm]{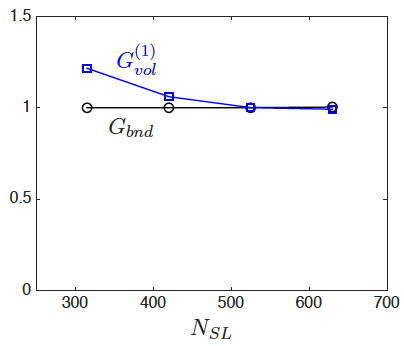}
}
\vspace{-0.2cm}
\caption{
Influence of mesh size: convergence with $N_{SL}$, the number of vertices across the shear layer (i.e. along the line $\{ -W/2 \leq x \leq W/2, y=D/2 \}$).
Circles: harmonic gain for boundary forcing on $\Gamma_f$;
squares: first optimal gain for volume forcing in $\dom$.
Gains are normalised by their values obtained on the reference mesh M$_3$ ($N_{SL}$=525, see table~\ref{tab:meshes}).
Forcing frequency $\omega_1$, amplitude $v'=0.075$~m/s.
} 
\label{fig:cvrg_msh}
\end{figure}

Convergence is reported in figure~\ref{fig:cvrg_msh} for two selected quantities:
harmonic gain $G_{bnd}$ for boundary forcing $\widetilde\ff_1$ on $\Gamma_f$, and 
first optimal gain $G_{vol}^{(1)}$ for volume forcing in $\dom$ 
(both at frequency $\omega_1$, around the LES mean flow at $v'=0.075$~m/s).
While $G_{bnd}$ is already well converged on the relatively coarser mesh M$_1$ ($0.3\%$ variation between M$_1$ and M$_4$), 
$G_{vol}^{(1)}$ requires the finer mesh M3 for a satisfactory convergence ($0.9\%$ variation between M$_3$ and M$_4$).
As mentioned in \S~\ref{sec:Numerical}, mesh M$_3$ is therefore used throughout the paper.

\section*{Appendix B. Sensitivity of harmonic gain}

Recall the LNSE  (\ref{eq:LNSE}) or, equivalently, the definition of the resolvent operator (\ref{eq:resolv}):
\begin{align}
(i \omega + \LL) \uu &=
\ff
\quad 
\Leftrightarrow
\quad
\uu 
= (i \omega + \LL)^{-1}\ff
= \RR(\omega) \ff.
\label{eq:LNSE3}
\end{align}
For the sake of simplicity and generality,
 we drop tildes $\widetilde\cdot$ and omit the dependence on the mean flow $\UUm$.
We distinguish two cases:  harmonic forcing $\ff_{vol}$ applied in the volume,
\be
(i \omega + \LL) \uu_{vol} = \ff_{vol} \quad\mbox{in \dom},
\qquad
\uu_{vol} = \zzz  \quad\mbox{on } \Gamma_f, 
\label{eq:res_vol}
\ee
and harmonic forcing $\ff_{bnd}$  applied at a boundary,
\be
(i \omega + \LL) \uu_{bnd} = \zzz \quad\mbox{in \dom},
\qquad
\uu_{bnd} = \ff_{bnd}  \quad\mbox{on } \Gamma_f.
\label{eq:res_bnd}
\ee
We write in short those two problems as
\be 
\uu_{vol} = \RR_{vol} \ff_{vol}
\qquad \mbox{and} \qquad
\uu_{bnd} = \RR_{bnd} \ff_{bnd}.
\ee

For a given forcing $\ff$, a variation of the  NS operator  $\LL \rightarrow \LL+\bdelta\LL$ induces a variation of the response $\uu \rightarrow \uu+\bdelta\uu$.
Substituting into (\ref{eq:res_vol})-(\ref{eq:res_bnd}), expanding and keeping only zeroth- and first-order terms yields  for volume forcing:
\be
(i \omega + \LL) \uu_{vol} 
+ (i \omega + \LL)\bdelta\uu_{vol} 
+ \bdelta\LL\,\uu_{vol}
= \ff_{vol} \quad\mbox{in \dom},
\qquad
\uu_{vol}+\bdelta\uu_{vol} = \zzz  \quad\mbox{on } \Gamma_f,
\ee
and for boundary forcing:
\be
(i \omega + \LL) \uu_{bnd} 
+ (i \omega + \LL)\bdelta\uu_{bnd} 
+ \bdelta\LL\,\uu_{bnd}
= \zzz \quad\mbox{in \dom},
\qquad
\uu_{bnd} + \bdelta\uu_{bnd} = \ff_{bnd}  \quad\mbox{on } \Gamma_f. 
\ee
Upon subtracting (\ref{eq:res_vol}) and (\ref{eq:res_bnd}), respectively, both problems reduce to:
\be
(i \omega + \LL)\bdelta\uu = -\bdelta\LL\,\uu
\quad\mbox{in \dom},
\qquad
\bdelta\uu = \zzz  \quad\mbox{on } \Gamma_f. 
\ee
That is, in both cases (volume forcing and boundary forcing), 
the response variation $\bdelta\uu$ is solution of a \textit{volume resolvent} problem (volume forcing $-\bdelta\LL\,\uu$ and homogeneous boundary conditions):
\be
\bdelta\uu_{vol} 
= - \RR_{vol} \bdelta\LL_{vol} \uu_{vol}
\qquad \mbox{and} \qquad
\bdelta\uu_{bnd} 
= - \RR_{vol} \bdelta\LL_{bnd} \uu_{bnd}
\label{eq:du_vol_bnd}
\ee

We now proceed to find the gain variation induced by the variation of the NS operator.
The variation of the gain
\begin{align}	
G^2 = \dfrac{||\uu||^2}{||\ff||^2} 
= \dfrac{\ps{\uu}{\uu}}{\ps{\ff}{\ff}},
\label{eq:Gopt2}
\end{align}
reads at zeroth and first orders: 
\begin{align}
G^2+\delta (G^2) 
= \dfrac{\ps{\uu+\bdelta\uu}{\uu+\bdelta\uu}}{\ps{\ff}{\ff}} 
= \dfrac{\ps{\uu}{\uu}}{\ps{\ff}{\ff}}
+ 2 \mbox{Re}
\left\{
\dfrac{\ps{\uu}{\bdelta\uu}}{\ps{\ff}{\ff}} 
\right\},
\end{align}
i.e. after subtracting (\ref{eq:Gopt2}) and multiplying by $||\ff||^2$:
\begin{align}	
\delta (G^2) ||\ff||^2
= 2 \mbox{Re} 
\left\{ 
\ps{\uu}{\bdelta\uu}
\right\}.
\end{align}
Substituting the response variation $\bdelta\uu$ from (\ref{eq:du_vol_bnd}), and using the definition of an adjoint operator,
one obtains for volume forcing:
\begin{align}	
\delta (G_{vol}^2) ||\ff_{vol}||^2
&= 
2 \mbox{Re} \left\{\ps{\uu_{vol}}{-\RR_{vol} \bdelta\LL_{vol} \uu_{vol}}\right\}
\nonumber
\\
&=  -2 \mbox{Re} \left\{\ps{\RR_{vol}^\dag \uu_{vol}}{ \bdelta\LL_{vol} \uu_{vol}}\right\}
\nonumber
\\
&= -2 \mbox{Re} \left\{\ps{G_{vol}^2\ff_{vol}}{ \bdelta\LL_{vol} \uu_{vol}}\right\},
\label{eq:dG2_vol}
\end{align}
and for inlet forcing:
\begin{align}	
\delta (G_{bnd}^2) ||\ff_{bnd}||^2
&=
2 \mbox{Re} \left\{\ps{\uu_{bnd}}{-\RR_{vol} \bdelta\LL_{bnd}\uu_{bnd}}\right\}
\nonumber
\\
&= -2 \mbox{Re} \left\{\ps{\RR_{vol}^\dag \uu_{bnd}}{ \bdelta\LL_{bnd}\uu_{bnd}}\right\}
\nonumber
\\
&= -2 \mbox{Re} \left\{\ps{\uu^\dag}{ \bdelta\LL_{bnd}\uu_{bnd}}\right\}.
\label{eq:dG2_bnd}
\end{align}
In (\ref{eq:dG2_vol}) we have used the  relation 
$\RR_{vol}^\dag \uu_{vol} = G_{vol}^2\ff_{vol}$ (\cite{Brandt11, Boujo15a}).
In (\ref{eq:dG2_bnd}), however, we have introduced 
$\uu^\dag=\RR_{vol}^\dag \uu_{bnd}$ (defined in the domain), which is \textit{not} equal to $G_{bnd}^2\ff_{bnd}$ (defined on the boundary).
Note that one can choose a unit forcing, $||\ff||=1$, since the gain is linear.

\begin{figure}[] 
\centerline{
\includegraphics[height=5cm]{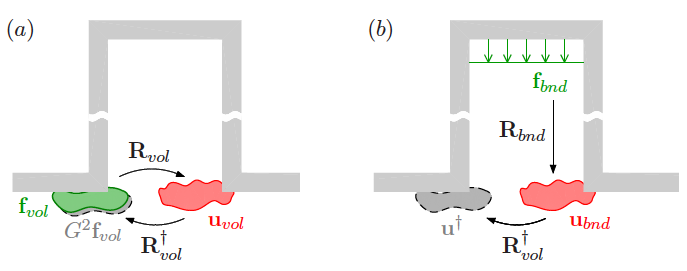}
}
\vspace{-0.2cm}
\caption{
$(a)$ Harmonic \textit{volume} forcing $\ff_{vol}$  and associated response $\uu_{vol}$, from which one  can compute the gain sensitivities (\ref{eq:sensit_volume_forcing_flow_modif}) and (\ref{eq:struct_sensit_volume_forcing}).
$(b)$~Harmonic \textit{boundary} forcing $\ff_{bnd}$  and associated response $\uu_{bnd}$.
The adjoint $\uu^\dag$ is needed to compute the gain sensitivities 
(\ref{eq:sensit_boundary_forcing_flow_modif})
and (\ref{eq:struct_sensit_boundary_forcing}).
} 
\label{fig:sketch_adjoint}
\end{figure}

As illustrated in figure~\ref{fig:sketch_adjoint},
knowing $\ff_{vol}$ and $\uu_{vol}$ is sufficient to compute the gain sensitivity in the case of \textit{volume} forcing;
the adjoint $\uu^\dag$ is necessary, however,
to compute the gain sensitivity in the case of \textit{boundary} forcing.

Expressions (\ref{eq:dG2_vol})-(\ref{eq:dG2_bnd}) allow one to easily compute gain variations $\delta(G^2)$ for any small-amplitude modification $\bdelta \LL$ of the NS operator, without solving explicitly for the modified response $\uu+\bdelta\uu$.
These expressions are general, but we can now make them more specific for two particular modifications $\bdelta\LL$ of interest.

First, when the mean flow is modified, $\UUm \rightarrow \UUm+\bdelta\UUm$, 
the NS operator variation reads
\be 
\bdelta\LL\,\uu = (\uu \bcdot \bnabla)\bdelta\UUm +
(\bdelta\UUm \bcdot \bnabla) \uu,
\ee 
and one obtains after a few manipulations:
\begin{align}	
\delta (G_{vol}^2) 
&= -2 G_{vol}^2 \mbox{Re} \left\{\ps{\ff_{vol}}{ (\uu_{vol} \bcdot \bnabla)\bdelta\UUm +
(\bdelta\UUm \bcdot \bnabla) \uu_{vol} }\right\},
\nonumber
\\
&= -2 G_{vol}^2 \mbox{Re} \left\{ \ps{\bnabla\uu_{vol}^H     \bcdot \ff_{vol} 
-\bnabla\ff_{vol} \bcdot \uu_{vol}^*}{\bdelta\UUm} \right\},
\\
\delta (G_{bnd}^2) 
&= -2 \mbox{Re} \left\{\ps{\uu^\dag}{ (\uu_{bnd} \bcdot \bnabla)\bdelta\UUm +
(\bdelta\UUm \bcdot \bnabla) \uu_{bnd} }\right\}
\nonumber
\\
&= -2 \mbox{Re} \left\{ \ps{\bnabla\uu_{bnd}^H \bcdot \uu^\dag 
-\bnabla\uu^\dag \bcdot \uu_{bnd}^*}{\bdelta\UUm} \right\},
\end{align}
hence the expressions (\ref{eq:sensit_volume_forcing_flow_modif})-(\ref{eq:sensit_boundary_forcing_flow_modif})
of the gain sensitivity $\bnabla_{\UUm} G^2$.

Second, for a feedback localised in $\xx=\xx_0$ in the form of a ``velocity-to-force'' coupling,
the NS operator variation reads
\be 
\bdelta\LL\,\uu = \CC(\xx)\uu =  \delta(\xx-\xx_0)\CC_0\uu
\ee
where $\delta(\xx)$ is the 2D delta Dirac function.
Expressions (\ref{eq:dG2_vol})-(\ref{eq:dG2_bnd}) therefore become
\begin{align}	
\delta (G_{vol}^2) 
&= -2 G_{vol}^2 \mbox{Re} \left\{\ps{\ff_{vol}}{ \delta(\xx-\xx_0)\CC_0 \uu_{vol}} \right\}
\nonumber
\\
&= -2 G_{vol}^2 \mbox{Re} \left\{ \ff_{vol}(\xx_0) \bcdot \CC_0 \uu_{vol}(\xx_0) \right\},
\\
\delta (G_{bnd}^2) 
&= -2 \mbox{Re} \left\{ \ps{\uu^\dag}{ \delta(\xx-\xx_0)\CC_0 \uu_{bnd}} \right\}
\nonumber
\\
&= -2 \mbox{Re} \left\{ \uu^\dag(\xx_0) \bcdot \CC_0 \uu_{bnd}(\xx_0) \right\}.
\end{align}
Choosing  the identity matrix for $\CC_0$ (i.e. a velocity sensed in the $x$ (resp. $y$) direction results in a force in the $x$ (resp. $y$)  direction only), one recovers the expressions (\ref{eq:struct_sensit_volume_forcing})-(\ref{eq:struct_sensit_boundary_forcing}) of the gain sensitivity $\bnabla_{\feedback} G^2$.

\section*{Appendix C. Influence of domain size}

A series of smaller LNSE domains are used to investigate which flow regions are important to capture the linear response to harmonic forcing.
Only the LNSE domain is modified: all linear response calculations are performed with the same LES mean flow.
The positions of the boundaries are varied as follows (see table~\ref{tab:domains}):
inlet ($x_1$) and outlet ($x_2$) in domains $D^{x}$,
cavity end ($y_2$) and lower channel wall ($y_1$) in domains $D^{y}$,
and all four boundaries in domains $D^{xy}$.
Note that the uniform harmonic forcing $\ff$ is applied on $\Gamma_f$, whose position $y_2$ varies in domains  $D^{y}$ and $D^{xy}$.

\begin{figure}[] 
\centerline{
\includegraphics[height=9.5cm]{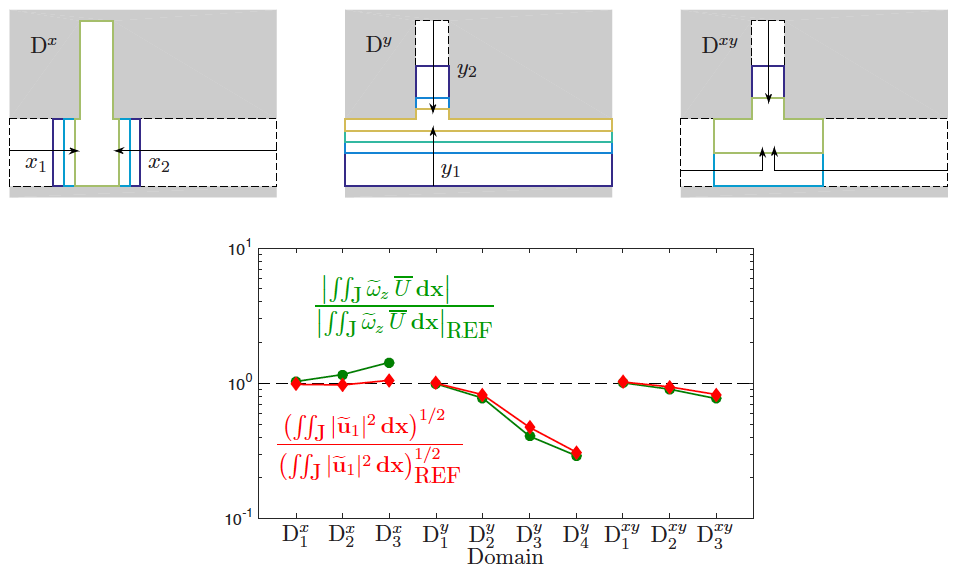}
}
\vspace{-0.2cm}
\caption{
Influence of domain size.
Linear harmonic response to boundary forcing on $\Gamma_f$.
Circles: kinetic energy (\ref{eq:gain_KE_subdom});  
diamonds: vertical component of the Coriolis force (\ref{eq:G_coriolis}).
Linear response values are normalised by the values obtained on the largest reference domain \dom (see table~\ref{tab:domains}).
Forcing frequency $\omega_1$, amplitude $v'=0.075$~m/s.
} 
\label{fig:cvrg_domain}
\end{figure}

The response measured in terms of kinetic energy (\ref{eq:gain_KE_subdom}) and vertical component of the Coriolis force (\ref{eq:G_coriolis}) is shown in figure~\ref{fig:cvrg_domain}, (normalised by values obtained on the largest reference domain \dom).
Note that domains $D^{y}$ and $D^{xy}$ have a smaller vertical extension than region \subdom, where the response is normally computed.

\begin{table}
  \begin{center}
\def~{\hphantom{0}}
  \begin{tabular}{c | c | ccc | cccc | ccc}
     & \textbf{REF (I)} &D$^x_1$&D$^x_2$&D$^x_3$&D$^y_1$&D$^y_2$&D$^y_3$&D$^y_4$&D$^{xy}_1$&D$^{xy}_2$&D$^{xy}_3$ \\ \hline 
$x_1$& \textbf{-80}     & -40   & -30   & -20   & -     & -     & -     & -     & -50      & -50      & -50\\
$x_2$& \textbf{165}     &  40   &  30   &  20   & -     & -     & -     & -     &  50      &  50      &  50\\
$y_2$& \textbf{121}     & -     & -     & -     &  80   &  50   &  40   &  40   &  80      &  50      &  50\\ 
$y_1$& \textbf{-31}     & -     & -     & -     & -     &   0   &  10   &  20   &  -       & -        &  0\\ 
  \end{tabular}
  \caption{Domains used for convergence study, with various locations of the inlet ($x_1$), outlet ($x_2$), cavity end ($y_2$) and lower channel wall ($y_1$). 
Dimensions in mm.
Only values different from the reference domain \dom are indicated.
See also figure \ref{fig:cvrg_domain}.
}
\label{tab:domains}
\end{center}
\end{table}

\bibliography{cavity_lnse_arxiv1}

\end{document}